\documentclass[10pt]{elsart}
\usepackage{amssymb,amsmath}
\usepackage{latexsym}
\usepackage{graphicx}
\usepackage{epsfig}
\journal{Chemical Physics}
\newcommand{\dt}[1]{\Delta\tau_{1#1}}
\begin{document}
\begin{frontmatter}
\title{Theory of ultrafast exciton dynamics in the Quantum Hall system}
\author{I. E.  Perakis\corauthref{cor}},
\corauth[cor]{Corresponding author} 
\ead{ilias.e.perakis@vanderbilt.edu}
\author{E. G. Kavousanaki}
\ead{elia@physics.uoc.gr} 
\address{Institute of Electronic Structure \& Laser, Foundation
for Research and Technology--Hellas
and
Department of Physics, University of Crete, 
P.O. Box 2208, 710 03, 
Heraklion, Crete, Greece} 

\date{\today}

\begin{abstract}
We discuss a  theory of 
the ultrafast non-linear optical response of excitons 
in the Quantum Hall system. 
Our theory focusses 
on the role 
of the low energy collective electronic excitations of the 
cold, strongly correlated,  two--dimensional electron gas (2DEG)
present in the
ground state. It 
 takes into account 
 ground state electron correlations and 
Pauli exchange  and interaction effects  
between the photoexcited excitons and the collective excitations. 
Our  formulation addresses 
both the initial coherent regime, 
where the dynamics is determined by exciton 
and 2DEG polarizations, 
and the subsequent incoherent regime, dominated by 
population dynamics.
We describe non-Markovian memory, dephasing, 
and correlation effects 
and a non-linear  exciton hybridization
due to the 
non-instantaneous interactions.
We identify the 
signature
of the coherent 
inter--Landau level magnetoplasmon (MP) 
in the temporal profile of the three--pulse 
time--integrated four--wave--mixing signal
in the initial coherent regime. 
\end{abstract}
\end{frontmatter}
%%%%%%%%%%%%%%%%%%%%%%%%%%%%%%%%%%%%%%%%%%%%%%%%%%%%%%%%

%%%%%%%%%%%%%%%%%%%%%%%%%%%%%%%%%%%%%%%%%%%%%%%%%

\section{Introduction}
In low dimensional geometries,  
collective many--electron effects can 
lead to new quantum phases that display novel transport and optical 
properties. 
One such example is the cold two--dimensional electron gas (2DEG) in 
modulation--doped quantum wells (MDQW), which 
provides an ideal ``laboratory'' for 
studying and manipulating charge
and  spin excitations  on the nanometer and femtosecond scale
\cite{QHE1,QHE2,QHE3}.
In the presence of a strong magnetic field pointing along the 
MDQW growth direction, the 2DEG 
is a strongly correlated quantum liquid 
whose ground state cannot be described 
by a non-interacting  Hartree--Fock state, 
unlike the ground state of undoped semiconductors 
\cite{QHE1,QHE2}. 
The magnetic field creates discrete Landau levels (LL),
which in the ground state are partially 
filled by  electrons introduced by the doping. 
These  LLs are  highly degenerate, 
and the states of each LL 
are strongly coupled by the Coulomb interaction.
The latter coupling  leads to 
collective charge and spin excitation modes, 
such as the intra--LL and inter--LL 
magnetoplasmons (MP), with dispersion governed by strong correlation effects
that lead to a magnetoroton minimum
\cite{QHE1,QHE2}. 
The 
LL degeneracy increases with magnetic field, 
and above a threshold value 
the ground state electrons only occupy the lowest LL 
(LL0) states;  all the higher LLs (LL1, LL2, $\cdots$) 
 are then empty in the ground state. 
The percentage of occupied states 
gives the LL filling factor, $\nu$. 
The ground state, well described by the 
Jastrow--like 
Laughlin wavefunction \cite{QHE1,QHE2}, depends 
on this filling factor.
Strong exchange Coulomb interactions stabilize a 
ferromagnetic ground state with polarized 
electron spins 
for  filling factors 
$\nu =1/m$, where $m$ is an integer
\cite{QHE2,ferro}, or for integer $\nu$.
For fractional values of $\nu$, the correlated 
ground state leads to the fractional QHE \cite{QHE1,QHE2,QHE3}. 

Recent time--resolved four--wave--mixing experiments 
shed new light into the dynamics
 of this strongly correlated system, and  opened 
 a new field of non-equilibrium Quantum Hall physics 
\cite{from-02-prl,from-99,kara,from-02-prb,per-03,per-dan-ssc,per-dan-pss,schu,from-02-ph,dani}. 
The interband optical absorption spectrum of the 2DEG 
is dominated 
by  LL exciton peaks 
\cite{from-02-prl,from-02-prb}. 
In the QHE system, 
the 2DEG correlations and collective electronic excitations 
dominate 
the dephasing of the photoexcited 
excitons (X) for low 
photoexcitation intensities
\cite{from-02-prl,from-99,kara,from-02-prb,per-03,per-dan-ssc,per-dan-pss,schu,from-02-ph,dani}. 
Of particular interest  
is the dynamics of the QHE system 
during 
time scales 
comparable to the 
characteristic time it takes the cold 2DEG ``bath'' system
to react to the introduction of  photoexcited
Xs.
The dephasing times of the two lowest 
LL excitons in the QHE system range from 
a few picoseconds (LL0) to a few hundreds of femtoseconds (LL1), 
while the 2DEG responds to the X--2DEG 
interactions 
within a time interval  comparable to 
the period of its  low energy  excitations.
The period of the 
lowest inter--LL 
magnetoplasmon (MP)
collective modes \cite{QHE1,QHE2}, discussed below, is 
 $T_{MP} =
2 \pi \hbar/\Omega_{M}$, where 
$\Omega_{M} \sim 15-20\: meV$ is the MP 
excitation energy. Therefore, 
  $T_{MP}$ 
is of the order  of a few 
hundreds of femtoseconds, 
longer than 
the duration of the  $\sim$100fs optical pulses
used to probe this system  
and tunable by changing  the magnetic field. 
Strong quantum kinetic effects in the ultrafast non-linear optical dynamics
 are expected in this regime, 
explored for the 
first time in \cite{from-02-prl,from-99,kara,from-02-prb,per-03,per-dan-ssc,per-dan-pss,schu,from-02-ph,dani}. 
In order to  study time scales   
comparable to the duration of the X--2DEG scattering events,  
well--established pictures such as the 
semiclassical Boltzmann picture of dephasing and relaxation
(which  assumes 
instantaneous X--2DEG scattering)  and  the Markov approximation
need to be revisited 
\cite{chemla-01,chemla-99,weg-00,haugbook,axt-04,rossi}. 

Transient
wave--mixing spectroscopy is  well suited for studying
strong correlation  and quantum coherent effects 
during time scales shorter than the dephasing times. 
Indeed,  
fluctuations beyond the Random Phase Approximation (RPA) can generate a
FWM signal with a distinct time dependence
\cite{chemla-01,chemla-99}.
For example, in undoped semiconductors, exciton--exciton 
(X--X) interactions 
dominate the two--pulse FWM signal
during negative time delays, 
where  the Pauli blocking effects vanish \cite{chemla-01,chemla-99}.
The time--dependent Hartree--Fock (HF) treatment of 
such interactions 
\cite{haugkoch} 
predicts an {\em asymmetric} 
FWM temporal profile, with  
a negative
time delay FWM signal decaying twice as fast as the  positive time delay
signal
\cite{chemla-01,chemla-99,shah-99}.
The observation of strong deviations 
from the asymmetric HF
temporal profile is  the  signature of strong 
 X--X correlations \cite{chemla-01,chemla-99}.
In this paper, our main goal is to describe theoretically the 
non-instantaneous many--body effects in the ultrafast 
non-linear optical response of the 
QHE system that are due to the correlated 2DEG.
We discuss ways in which 
ultrafast non-linear optical spectroscopy 
can be used to probe  the {\em coherent MP dynamics}, 
important for understanding and 
manipulating the coherent transport and optical properties of the 2DEG 
but not accessible  so far. 

The 
two--pulse wave mixing
spectroscopic studies 
of electronic dephasing in the 
QHE regime 
reported in \cite{from-02-prl,from-99,kara,from-02-prb,per-03,per-dan-ssc,per-dan-pss,schu,from-02-ph} 
focussed on the dynamics of the excitonic peaks 
for filling factors 
 $\nu<2$, 
where only LL0 states are populated in the ground state.  
The time and frequency dependence of the FWM spectra 
revealed new dynamical features 
that
 could not be explained within the 
RPA.
The most important observations 
were a strong optically--induced time--dependent hybridization 
of the LL0 and LL1 exciton peaks  governed by the 
inter--LL MP dynamics,  
a  non-Markovian dephasing 
of the LL1 exciton  peak when $\nu < 2$, and strong oscillations
as function of time delay of quantum kinetic origin.
The exciton hybridization 
manifests itself 
in a strong enhancement of the LL0
FWM signal  even in the case of  minor 
photoexcitation of the LL0 optical transitions.  
This non-resonant FWM signal displayed a 
symmetric time delay profile,
unlike for the LL1 signal and the predictions of the RPA.
Additionally, a strong dephasing of the LL1 resonance 
and an asymmetric LL1 linear absorption lineshape 
was observed when $\nu <2$, unlike for the LL0 resonance.
Finally, for equal photoexcitation of the LL0 
and LL1 transitions, 
the FWM signal at the LL1 energy was very minimal, 
while the  LL0 FWM resonance showed strong oscillations 
as function of time delay, despite the fact that 
a single LL0 peak dominated the FWM spectral profile.
In contrast to the MDQW, a comparable undoped quantum well 
did not display any of the above behavior: its FWM spectrum 
was consistent, to first approximation,  
with the predictions of a three level system
consisting of the ground state, LL0, and LL1 levels. 
Importantly, the strength of the new MDQW  features
diminished as 
the density of photoexcited carriers approached the 
density of the ground state 2DEG.
In this case, 
the Coulomb scattering among the photoexcited carriers
dominates over the X--2DEG interactions 
and can even change the dephasing times.
More recent experiments by the Chemla group \cite{dani}  
showed that the initial coherent regime, which lasts for a few 
picoseconds, is followed by an incoherent regime that, for low 
photoexcitation intensity, 
builds up slowly over a time interval of several picoseconds. 

To interpret experiments such as the above, 
one should note some important differences 
between the MDQW and QHE system and the undoped QWs 
 studied so far. 
In addition to the  ground state electron correlations 
in the QHE system, important differences 
in the optical properties between the doped and undoped QWs 
 arise from the different nature of the 
low energy electronic excitation spectrum in the two systems. 
In undoped semiconductors, the lowest
electronic excitations of the ground state electrons are the high
energy interband $e$--$h$ pair excitations, which can adjust
almost instantaneously to the dynamics of the photoexcited
carriers. The latter then behave as quasi--particles with mutual
interactions, while the ground state can be considered as rigid:
the many--body nature of the system only affects the different
parameters associated with the band structure and the dielectric
screening.

In the case of the 2DEG in the QHE regime 
one should distinguish 
between the excitations of two subsystems:
(i) the QW interband excitations 
(with the 2DEG at rest), which consist of
1 $e$--$h$, 2 $e$--$h$, $\cdots$ pairs  created in the LLs, and (ii) 
the 2DEG excitations (with unexcited
QW and full valence band), 
e.g. the  1--MP, 2--MP, $\cdots$ 
and incoherent pair excitation 
 states.
The ensemble of
states that determine the non-linear optical spectra
can be thought of as consisting of 
$\ell$ $e$--$h$ pairs 
 and $n$ 2DEG excitations. 
One can then draw an analogy between the
X--MP effects studied here and  the X--phonon 
effects in the ultrafast 
 response of  undoped semiconductors
that were extensively studied in the past 
\cite{weg-00,haugbook,rossi,axt98,axt96,schilp-94,ost-99,vu-00}.
In both cases, the ultrafast 
non-linear optical dynamics is governed by
the interactions between photoexcited Xs and a
collective excitation. 
However, there
are some important differences. 
In the QHE system, 
the collective excitations are 
electronic in nature, and 
are described 
by similar electronic operators
as the photoexcited carriers.  
They are therefore subject to Pauli correlations 
with the photoexcited Xs, 
while the 
ground state 
electrons
are strongly correlated and 
sustain low--energy excitations. 
On the other hand, in the undoped system, 
the electronic operators
commute
with the collective 
excitation (phonon) operators, 
while the ground state correlations can be 
neglected.

The role of the Coulomb and exciton--phonon correlations
in undoped semiconductors  
have been studied in the past using different theoretical approaches 
\cite{haugbook,rossi,axt98,axt96,ost-98,schaf-book,schaf-96,loven-02,bind,mukamel-book,per-00,shah00}.
A widely used approach
is the ``dynamics--controlled truncation
scheme'' (DCTS) \cite{axt-04,axt98,axt96}, 
which truncates the 
hierarchy of density matrices 
generated  by the interactions  based on the fact that,
in the undoped system, all Coulomb
interactions occur between photoexcited $e$--$h$ pairs and are thus
dynamically generated by the optical excitation, 
treated with an expansion in terms of the optical field. 
The DCTS  also assumes the absence of free carriers in the
ground state \cite{axt98}, a condition that does not hold in
the MDQWs. 
The almost unexplored dynamics of strongly
correlated systems whose ground state electrons interact
unadiabatically with the photo--excited $e$--$h$ pairs raises very
fundamental questions. In MDQWs, the direct exciton--exciton
interactions are screened, and the non-linear response is mainly
determined  by the Fermi sea excitations
\cite{per-00,bre-95,barad-94,per-94,per-96-b,prim-00,shah-00-a}.
In the QHE regime, 
the presence of collective low
energy electronic excitations and the resulting non-Markovian dynamics 
and memory effects, as well as 
the strongly correlated ground state, raise
formidable theoretical difficulties for describing the non-linear optical 
dynamics.
For example, 
standard diagrammatic expansions and 
DCTS factorizations that assume a 
Hartree--Fock reference state break down.
In addition,
the theory must address the quantum effects 
due to the Pauli correlations between the 
collective electronic  excitations and the photoexcited carriers. 

In the first part of this paper we discuss a theoretical formulation 
that describes the ultrafast 
third--order 
non-linear optical response of the QHE system
at zero temperature  
and addresses both the 
coherent and incoherent regimes. 
To our knowledge, this is the first theory 
that addresses this non-equilibrium many--body problem. 
Our approach is based on the projection of the exciton 
states and the separation of the uncorrelated
contributions to the third--order non-linear optical response   
from the contributions due to correlations among the 
interband and intraband  elementary 
excitations.
We also introduce a basis of correlated states, 
generated by using the Lanczos iterative approach \cite{haydock}, 
which is useful for calculating the 2DEG correlations
\cite{QHE1,QHE2} 
as well as 
for deriving a simple average polarization model 
\cite{chemla-99,schaf-96,shah00}
for treating the 
ultrafast dynamics. 
We neglect phonons since the optical 
phonon frequency is much larger than the inter--LL 
splitting.
In the second part of the paper,  we focus on the coherent regime 
and discuss the manifestations of the 
MP coherence 
in the three--pulse
time--integrated FWM signal,
while neglecting the incoherent effects. 
This calculation is based on an average polarization 
model         derived from the general theory by 
projecting out the first few correlated Lanczos states. 

Although  our theory can be used to treat the general case, 
of particular interest here is the 
spin--$\uparrow$ polarized 
LL0 2DEG, realized 
for filling factors $\nu = 1/m$, where $m$ is an integer, 
or integer $\nu$. We consider the case of 
photoexcitation with 
right--circularly polarized optical pulses, 
which excite spin--$\downarrow$ electrons.
For simplicity we only consider the LL0 and LL1 electron and hole states. 
We show that  
the MP effects can be separated 
from the corresponding Pauli blocking and exciton--exciton  
interaction signals based on the temporal profile 
of the FWM  signal.
Throughout the paper we point out the analogies between the 
theoretical formulation of the 
X--MP effects presented here 
and the  DCTS \cite{axt98,axt96} 
and correlation expansion \cite{rossi}
approaches for treating 
X--phonon effects in undoped semiconductors. 
Similar to the DCTS \cite{axt96}, we use an expansion 
in terms of the optical field in order to eliminate the 
number of independent dynamical variables that need to be considered. 
The  X--2DEG correlations do not allow the complete factorization 
of the intraband density matrix into products of interband coherences. 
Here we separate out the correlated contributions, which lead 
to the incoherent effects, without assuming a Hartree--Fock ground state, 
as in the case of the DCTS. In undoped semiconductors,
our approach reduces  to the DCTS if phonons are included. 

In Section \ref{setup} we set up the general problem 
and in Section \ref{MP-MEX} 
we discuss the nature of the exciton and magnetoplasmon 
states that determine the  optical spectra.
We also present some useful relations for describing the Pauli 
exchange quantum effects  and the interaction effects between 
Xs and MPs. 
In Section \ref{nlo}
we present the equations of motion for the
non-linear polarizations and photoexcited carrier populations
and identify  the  contributions due to the many--body 
correlations.
In Section \ref{correl}
 we present a decoupling scheme for treating 
the interaction effects, which is motivated by 
a decomposition of the photoexcited many--body states 
that separates out the uncorrelated and excitonic 
contributions from the correlated and incoherent contributions. 
We use this approach to devise a factorization scheme  
and identify the intraband and interband correlated contributions
to the density matrix.
We also discuss the linear absorption spectra.
In Section \ref{XX}
we briefly discuss the coherent X--X interaction 
and scattering effects and derive an average polarization 
model \cite{chemla-99,schaf-96,shah00} for treating such correlation 
effects by using a
basis of interacting X--X Lanczos states \cite{haydock}.
In Section \ref{GPA}
we derive from the general equations  a generalized average 
polarization model 
that treats the effects  of the magnetoplasmon 
dynamics. Finally, in Section \ref{results},  
we use the above model 
to calculate the three--pulse time--integrated FWM signal  
and identify the signatures of the MP dynamics in the coherent regime
by assuming that 
incoherent effects can be neglected for such short time scales. 
We end with the conclusions.

\section{Hamiltonian }

\label{setup}

To describe the optical response of the QHE system, we adopt   
the standard two--band Hamiltonian of
interacting electrons and holes 
coupled by an optical field $E(t)$
and subject to a magnetic field, 
treated within the Landau gauge 
${\bf A}=(0, Bx , 0)$, that splits
the conduction and valence bands into 
discrete electron ($e$) and hole
($h$) LLs \cite{haugkoch,schaf-book}:
\begin{equation} \label{H-tot}
H_{{\rm tot}}(t) = H - \mu E(t) {\hat X}^{\dag}-  
\mu E^\ast(t)  {\hat X},
\end{equation}
where  
$\hat{X}$ 
is the interband optical transition operator, 
discussed further in the following section,    
$\mu$ is the interband transition matrix element
and $N =L^2/2 \pi l^2$ is the LL degeneracy,
where $l=(\hbar c/e B)^{1/2}$ is the magnetic length (Larmor radius) 
and $L$  the system size \cite{shah00,macd-85-1}.
$H$ is the ``bare'' semiconductor many--body Hamiltonian 
($\hbar=1$ from now on)  
\cite{haugkoch,schaf-book,shah00,kara}, 
\begin{equation} \label{H}
H =  \sum_{ k n \sigma} [E_g + \Omega_c^c (n + 1/2)] \, 
 {\hat
e}^{\dag}_{kn\sigma} {\hat e}_{kn\sigma} 
+ \sum_{ kn \sigma}  \Omega_c^v(n + 1/2)\,
{\hat h}_{kn \sigma}^{\dag} {\hat
h}_{kn\sigma}
 + H_{int}, 
\end{equation}
where $E_{g}$ is the bandgap and 
$\Omega^{i}_c =  e B/m_i$, $i=c,v$, 
are the electron and hole cyclotron energies
that determine the Landau level spacings, which 
are inversely proportional to the 
carrier masses and  closely spaced 
 in the valence band. 
In the above Hamiltonian, 
${\hat e}^{\dag}_{kn\sigma}$ is the creation operator of the 
spin $\sigma$
LL$n$ conduction band
electron state 
\begin{equation} 
\label{wave}
\psi_{kn }({\bf r})
 = \frac{e^{i k y}}{\sqrt{L}}
 \phi_n(x-x_k) 
\end{equation} 
characterized 
 by the momentum $k=x_k/ l^2$, where  
$x_k$
is the x coordinate of the cyclotron orbit center. 
In the above equation, 
$\phi_n$ is the $n$--th eigenfunction of the 1D
harmonic oscillator, with energy equal to the cyclotron 
energy.
Similarly, $\hat{h}^{\dag}_{kn\sigma}$ creates the 
LL$n$ valence band
hole state $\bar{\psi}$, which in the ideal two--dimensional 
system is related to the conduction electron wavefunction
by   $\bar{\psi}_{kn}=\psi^*_{-kn}.$
We only consider heavy hole states,  
with angular momentum $J=3/2$ 
 and $m=3/2$ ($\sigma=\uparrow$) 
or $m=-3/2$ ($\sigma=\downarrow$).
The above relation  
between the conduction electron and valence hole wavefunctions 
in the ideal two--dimensional system
leads to an electron--hole symmetry 
that strongly affects the non-linear optical properties \cite{shah00,apalkov}. 
In the realistic system, this symmetry is lifted due to the lateral 
confinement, the different band offsets 
and confinement between the electrons and the hole, 
the valence band mixing, etc. 

The Hamiltonian 
$ H_{int}$ 
describes the {\em e--e}, {\em e--h}, and {\em h--h}
Coulomb interactions, 
\begin{align}
H_{int}=\frac{1}{2}\sum_{\alpha_1\alpha_2\alpha_3\alpha_4}  
\Bigl[&
v_{\alpha_1\alpha_2,\alpha_3\alpha_4}^{ee}
\hat{e}_{\alpha_3}^{\dag}\hat{e}_{\alpha_1}^{\dag}\hat{e}_{\alpha_2}
\hat{e}_{\alpha_4} 
+ v_{\alpha_1\alpha_2,\alpha_3\alpha_4}^{hh}
\hat{h}_{\alpha_3}^{\dag}\hat{h}_{\alpha_1}^{\dag}
\hat{h}_{\alpha_2}\hat{h}_{\alpha_4} \nonumber \\
&- v_{\alpha_1\alpha_2,\alpha_3\alpha_4}^{eh}
\hat{h}_{\alpha_3}^{\dag}\hat{e}_{\alpha_1}^{\dag}\hat{e}_{\alpha_2}
\hat{h}_{\alpha_4}
- v_{\alpha_1\alpha_2,\alpha_3\alpha_4}^{he}
\hat{e}_{\alpha_3}^{\dag}\hat{h}_{\alpha_1}^{\dag}\hat{h}_{\alpha_2}
\hat{e}_{\alpha_4} \Bigr], \label{Hint}
\end{align}
where $\alpha = (k,n,\sigma)$. 
In the ideal two--dimensional 
system, 
the Coulomb interaction matrix elements
$v_{\alpha_1\alpha_2,\alpha_3\alpha_4}^{ij}$ (with $i,j=e,h$) are
given by
\begin{equation}
\label{V-alpha} v_{\alpha_1\alpha_2,\alpha_3\alpha_4}^{ij}= \int 
\frac{d{\bf q}}{(2\pi)^2}v_{q}F_{\alpha_1\alpha_2}^{i}({\bf q})
F_{\alpha_3\alpha_4}^{j}(-{\bf q}),
\end{equation}
where $v_{q}=2 \pi e^2/q$ is the 2D Coulomb potential, and  
\cite{macd-85-1}
\begin{equation} 
\label{Fe}
F_{\alpha_1\alpha_2}^{e}({\bf q})= 
\varphi_{n_1n_2}({\bf q})
e^{iq_x(k_1+k_2)l^2/2}\delta_{k_1,k_2+q_y}
\delta_{\sigma_1,\sigma_2} \ , \ 
 F_{\alpha_1\alpha_2}^{h}({\bf q})= 
F_{-\alpha_2,-\alpha_1}^e({\bf q}), 
\end{equation} 
where $-\alpha = (-k,n,\sigma)$ and 
\begin{equation}
\label{phi} 
 \varphi_{mn}({\bf q})=
\frac{n!}{m!}\biggl[\frac{(-q_y + i q_x)l}{\sqrt{2}}\biggr]^{m-n}
L_{n}^{m-n}\biggl(\frac{q^2l^2}{2}\biggr)
e^{-q^2 l^2/4}, 
\end{equation}
for $m \ge n$ and 
$\varphi_{mn}({\bf q})=
\varphi^*_{nm}(-{\bf q})$  
for $m < n$, 
where $L_n^{m-n}$ is the generalized Laguerre polynomial.
From now on we measure all energies with respect
to the ground state energy and thus have that 
$H | G \rangle =0$.

\section{2DEG Magnetoexcitons and Magnetoplasmons} 

\label{MP-MEX} 

In this section we briefly discuss the 
magnetoexciton (X) and magnetoplasmon (MP)  
excitations that govern the
ultrafast non-linear optical dynamics. 
From now on  we restrict to the case of photoexcitation 
with right--circularly polarized light, 
which excites spin--$\downarrow$ $e$--$h$ pairs.
We start with the 
dipole transition operator $\hat{X}^\dag$, 
which can be expanded in terms of the 
exciton creation operators $\hat{X}_i^\dag$ that create the allowed 
optical transitions 
for given LLs.
We introduce the creation operators 
of LL$m \to$ LL$n$ magnetoexcitons
with total momentum ${\bf q}$, 
\begin{equation}
\label{Xq}  
\hat{X}^\dag_{{\bf q} nm} 
= \frac{1}{\sqrt{N}} 
\sum_k e^{ i k  q_x l^2} \ 
\hat{e}^\dag_{k+ q_y/2,n,\downarrow} \
\hat{h}^\dag_{-k + q_y/2,m,\downarrow}. 
\end{equation} 
In the absence of disorder, momentum is conserved and  
 only  ${\bf q}=0$ 
excitons are photoexcited directly. 
Furthermore, in the ideal system, 
the only allowed optical transitions
correspond to $m=n$. 
We then have that 
\begin{equation}
\label{Xn} 
\hat{X}^\dag 
= \sqrt{N} \sum_n \hat{X}^\dag_n \ , \ 
\hat{X}_n^\dag =  \hat{X}^\dag_{0 nn}.
\end{equation} 
We note however that
the disorder present in the MDQW system can relax the 
momentum conservation condition \cite{aron-rev,aron-92,raman,marmorkos},
thus mixing X states with different momenta, 
while the valence band mixing couples the 
$n \ne m$ valence hole states and magnetoexcitons 
\cite{aron-rev,aron-92,sham-85}.
For this reason, in the formulation of the non-linear optical response 
we first consider all allowed
magnetoexciton transitions, $\hat{X}^\dag= \sqrt{N}\sum_i \hat{X}_i^\dag$, 
before presenting results 
specific to the ideal 2D system. 

The states $|X_i \rangle 
= \hat{X}^\dag_{i} | G \rangle$  
are the magnetoexciton states in the 2DEG system.
The difference from undoped semiconductors is that here 
the exciton operators $\hat{X}^\dag_{i}$
act on the 
strongly correlated state
$| G\rangle$, which  is  the ground eigenstate 
of  the many--body Hamiltonian $H$ that describes the 
correlated 2DEG at rest.  
Therefore, the 
 states $| X_i \rangle$ 
are strongly  correlated, 
similar to the magnetoplasmon states discussed below. 
The following orthogonality relation holds: 
\begin{equation} 
\label{X-ortho} 
\langle X_{i'} | X_i \rangle = ( 1 - \nu_{i}) 
\delta_{ii'}, 
\end{equation} 
where, as discussed below, $\nu_{i}$ 
is related to the LL filling factor.
For the 
magnetic fields of interest,
only the electron LL0 is partially filled in the  ground
state  with the 2DEG at rest. 
On the other hand,  
all  the hole LL states are empty (full valence band),
similar to the undoped system. 

We now turn to the magnetoplasmon modes, 
which dominate over quasi-electron -- quasi-hole pair excitations 
for  momenta $q < 1/l$. 
A LL$m\to$LL$n$ MP may be thought of  as an $e$--$h$ pair, or exciton, 
formed by an electron in   LL$n$  and a hole in the LL$m$ 2DEG.
The creation operator 
of this MP is given to first approximation  
by the LL$m\to$LL$n$ 
contribution to the collective density operator
\cite{QHE1,QHE2,oji,kallin-84}:  
\begin{equation}
\label{MP}
\hat{\rho}_{{\bf q}nm\sigma}^{e}  
= \frac{1}{\sqrt{N}} \sum_{k} 
e^{i q_x k l^2}\ 
 \hat{e}^\dag_{k+ q_y/2,n,\sigma} \
\hat{e}_{k- q_y/2,m,\sigma}.  
\end{equation} 
The analogy between the above MP and magnetoexciton 
creation operators is clear. 
It is convenient to also introduce  a similar collective operator 
for the hole states \cite{apalkov}, 
\begin{equation}
\label{rho-h} 
\hat{\rho}_{{\bf q}nm\sigma}^{h}  
= \frac{1}{\sqrt{N}} \sum_{k} 
e^{i q_x k l^2} 
 \hat{h}^\dag_{-k+ q_y/2,n,\sigma} \
\hat{h}_{-k- q_y/2,m,\sigma},  
\end{equation}   
and note the  relation 
\begin{equation} 
\label{rho-dag} 
\hat{\rho}^{i \dag}_{{\bf q} nm\sigma} 
= \hat{\rho}^{i}_{-{\bf q}mn\sigma} 
\ , \ i=e,h
\end{equation} 
that relates the creation and annihilation operators. 
Here we focus on photoexcitation of  the LL0 and LL1 optical
transitions only, which
are dynamically coupled by the 
LL0 $\to$ LL1 inter--LL MPs
 \cite{from-02-prl,kara,from-02-prb,per-03,per-dan-ssc,per-dan-pss}.
Theese  MPs 
are the lowest--energy neutral charge  excitations 
of the $\nu=1$ QHE ferromagnet, where the intra--LL charge excitations 
are suppressed since all spin--$\uparrow$ LL0 states 
are occupied in the ground state  \cite{ferro}. 

Similar to Feynmann's theory of the 
collective charge excitation
spectrum of liquid helium 
\cite{QHE1,QHE2,oji,feyn}, 
a good variational approximation of the MP eigenstates 
of the Hamiltonian $H$ 
is 
given by the state (single--mode approximation)
\cite{QHE1,QHE2}  
\begin{equation} 
\label{MP-state1} 
 | M_{{\bf q}} \rangle
= \sum_{\sigma n m } C_{nm\sigma}({\bf q}) 
\phi_{nm}({\bf q}) 
 \hat{\rho}_{{\bf q}nm \sigma}^{e} | G \rangle, 
\end{equation} 
where $C_{nm\sigma}({\bf q})$ are variational parameters. 
The mixing of the different  LL
states, which is due to the interactions,  
is suppressed in the strong magnetic field limit  \cite{QHE2}
by a factor $\propto B^{-1/2}$  
since the 
characteristic 2DEG Coulomb interaction energy   
$ e^2/l \propto \sqrt{B}$  
is smaller than the energy 
separation between the electron LLs, 
$\Omega_c^c \propto B$.
Even though in the realistic system 
$ e^2/l \sim \Omega_c^c$, 
calculations have shown that the  LL mixing 
does not change qualitatively
the MP 
and fractional QHE properties
\cite{macd-85-1,oji}.

We now turn to the Pauli exchange effects 
between the Xs.
As already known 
from undoped semiconductors, 
the latter 
are described by the 
deviation of the 
commutator of the X operators from bosonic behavior due to the 
underlying Fermi statistics. 
By using the second quantization 
expressions of  $\hat{X}$  and $\hat{\rho}$ 
and Eq. (\ref{X-ortho}) we obtain that 
\begin{equation} 
\label{commut-XX-q}
\left[ \hat{X}_{{\bf q}nm}, \hat{X}^{\dag}_{n'} \right] 
= \langle X_{{\bf q}nm} | X_n' \rangle 
- \frac{\delta_{n'm}}{\sqrt{N}} 
\Delta \hat{\rho}_{-{\bf q}n'n\downarrow}^{e}  
- \frac{\delta_{nn'}}{\sqrt{N}} 
\hat{\rho}_{-{\bf q}n'm\downarrow}^{h},
\end{equation} 
where
$\Delta \hat{\rho}_{{\bf q}nm\downarrow}^{e}  = 
\hat{\rho}_{{\bf q}nm\downarrow}^{e}  
- \langle G | \hat{\rho}_{{\bf q}nm\downarrow}^{e}| G \rangle$
and, for the strongly correlated 2DEG ground state, 
the latter expectation value vanishes for nonzero momentum 
(and for $n,m \ne 0$ in the strong magnetic field limit). 
More generally, we have that 
\begin{equation} 
\label{commut-XX-gen} 
[{\hat X}_i, {\hat X}_{i'}^{\dag}] = \delta_{ii'}
\left( 1 - \nu_i\right)  - \Delta \hat{\nu}_{ii'}
\end{equation} 
where the first term on the rhs describes the 
ground state contribution.
In the  case of zero--momentum 
excitons,
\begin{equation} 
\label{commut-XX} 
[{\hat X}_n, {\hat X}_{n'}^{\dag}] = \delta_{nn'}
\left( 1 - \nu_n  - \Delta \hat{\nu}_{n}\right), 
\end{equation} 
where
\begin{equation}
\nu_n= \frac{1}{N} \sum_{k} \langle G| 
{\hat e}_{kn\downarrow}^{\dag} {\hat e}_{kn\downarrow} | G\rangle 
\label{fill} 
\end{equation} 
is the ground state filling factor of the LL$n$ 
spin--$\downarrow$ 
2DEG electron states 
and 
\begin{equation}
\label{Dnu}
\Delta \hat{\nu}_n
= \frac{1}{N} 
\sum_{ k} \left( {\hat
h}_{- kn \downarrow}^{\dag}  {\hat h}_{- kn \downarrow} + 
{\hat e}_{kn\downarrow}^{\dag} {\hat e}_{k n\downarrow} 
- \langle G | {\hat e}_{kn\downarrow}^{\dag} {\hat e}_{k n\downarrow}
| G \rangle 
\right)
\end{equation}
describes the change in the LL$n$ filling factor due to the photoexcited 
electron and hole populations. 

Xs and MPs are made of electrons, and therefore 
the X--MP Pauli exchange 
effects  must also be considered. Similar to the case of Xs, 
these are  
described 
by the commutators 
\begin{equation} 
[ \hat{\rho}^e_{{\bf q}nm\sigma}, \hat{X}^\dag_{n'} ]
= \frac{1}{\sqrt{N}} \delta_{mn'} \delta_{\sigma\downarrow} 
\hat{X}^\dag_{{\bf q}nn'} \ , \ 
[ \hat{\rho}^h_{{\bf q}nm\sigma}, \hat{X}^\dag_{n'} ]
= \frac{1}{\sqrt{N}} \delta_{mn'} \delta_{\sigma\downarrow} 
\hat{X}^\dag_{{\bf q}n'n},
\label{commut-rho-X} 
\end{equation} 
obtained by using the above second quantization 
expressions for $\hat{\rho}$ and $\hat{X}$. 
In the case of a spin--polarized ground state 2DEG
of spin--$\uparrow$ 
electrons,  the MP and X operators commute 
since right--circularly polarized light 
creates spin--$\downarrow$ electrons.

In addition to the Pauli exchange effects, 
the optical properties are strongly affected by 
the interactions between 
the photoexcited excitons
$\hat{X}_{i}$ and the 2DEG carriers.  
We can describe such X--2DEG interactions, 
which scatter the X into X+MP final states,  
by considering the action of the Hamiltonian $H$ 
on $| X_i \rangle$ \cite{kara}:
\begin{equation}
H |X_i \rangle 
= \Omega_i |X_i \rangle - ( 1 - \nu_i) \sum_{i' \neq i} V_{i'i}
|X_{i'} \rangle + |Y_i \rangle. \label{HonX}
\end{equation}
The above equation 
defines the state $| Y_i \rangle$ by the requirement 
that it is orthogonal to all exciton states, 
$\langle X_j | Y_i \rangle=0$, and therefore 
describes an excited 2DEG configuration. 
Such configurations are denoted as 
2DEG$^\ast$ from now on. 
The above orthogonality requirement, as well as the orthogonality among the 
exciton states,  
gives 
\begin{equation}
\label{Xen} \Omega_i = \frac{\langle X_{i}| H | X_i \rangle}{
\langle X_{i} | X_i \rangle}, 
\end{equation}
the $X_i$ energy, and 
\begin{equation}
\label{V} 
V_{ii'} = - \frac{\langle  X_{i}| H | X_{i'} \rangle}{
(1 - \nu_i) ( 1 - \nu_{i'})}= 
V_{i'i}^{*},
\end{equation}
the static Coulomb--induced coupling of the different 
LL Xs.
Based on the above we introduce the operator 
\begin{equation}
\label{Yop} {\hat Y}_i = [{\hat X}_i,H] - \Omega_i {\hat X}_i +
( 1 - \nu_i) \sum_{i' \neq i} V_{ii'} {\hat X}_{i'},
\end{equation}
that describes the interactions between
$X_i$ and all the other carriers,   
X's, MPs, or ground state 2DEG.

An explicit expression for the operator ${\hat Y}_n$
was obtained in \cite{kara} 
by calculating the commutator 
$[\hat{X}_{n}, H ]$
using the 
Hamiltonian $H_{int}$ 
discussed in section \ref{setup}.
By only retaining 
contributions  from the photoexcited 
LLs (LL0 and LL1) we  obtain that 
\begin{align}
\label{XH} 
[\hat{X}_{n}, H ] =&
\left[ E_g + (n + 1/2) \left(\Omega_c^c + \Omega_c^v \right) 
\right]  \hat{X}_{n}\nonumber\\
& - ( 1 - \nu_n) \sum_{n'}  \hat{X}_{n'} \, 
\int
\frac{d{\bf q}}{(2\pi)^2} \, v_{q} \, |\phi_{nn'}(q)|^2
+ (\delta_{n1}-\delta_{n0})  \hat{Y}_{int}
\end{align} 
where the first two terms 
give  X energies and Coulomb--induced
couplings similar to the undoped system \cite{shah00,staff90},
while the interaction contributions  are described by the operator
\begin{equation} 
\label{Yint}
\hat{Y}_{int} 
=  \frac{1}{2 \pi l^2 \sqrt{N}} \,
\sum_{{\bf q}} \ \upsilon_q 
\hat{\rho}_{{\bf q}}
\left[ 
\phi_{10}(-{\bf q}) 
\hat{X}_{{\bf q} 01}  - 
\phi_{01}(-{\bf q}) 
\hat{X}_{{\bf q} 10} \right],
\end{equation} 
where we defined for simplicity 
\begin{equation} 
\label{ro} 
\hat{\rho}_{{\bf q}} = 
\sum_{ mm'\sigma} 
\phi_{mm'}({\bf q}) 
\left( \hat{\rho}^{e}_{{\bf q}mm'\sigma}
-\hat{\rho}^{h}_{{\bf q}m'm\sigma} \right). 
\end{equation} 
The operator $\hat{Y}$ can  be obtained 
from Eq. (\ref{Yop}) 
by subtracting from the above expression for $\hat{Y}_{int}$ 
the contributions to the X energies and couplings, 
Eqs. (\ref{Xen}) and (\ref{V}),  
which are $\propto \langle X_n | \hat{Y}_{int}^\dag | G \rangle$.
If we restrict to the first two LLs 
we have the simple property  
\begin{equation} 
\label{Ysymm}
\hat{Y}_1 = - \hat{Y}_0 = \hat{Y}. 
\end{equation} 

In the undoped system, 
we have that 
$\hat{Y}_{int}^\dag | G \rangle =0$
and 
$\hat{Y} = \hat{Y}_{int}$ describes the X--X interaction effects. 
In  the doped system, 
 $\hat{Y}_{int}$ renormalizes 
the X energies and couplings
due to the X interactions with the ground state 2DEG.
In addition to 
X--X interactions, 
 $\hat{Y}_{int}$ 
describes X+MP scattering effects. This can be seen 
from Eqs. (\ref{Yint}) and (\ref{ro})   
by recalling that the operators 
$ \hat{\rho}^{e}_{{\bf q}mm'\sigma}$
create and annihilate the MPs.
To make the analogy to the case of phonons in undoped semiconductors 
\cite{rossi,axt96}, 
 $\langle \hat{Y} \rangle$ describes the contribution of the MP--assisted 
interband 
density matrices and the X--X interactions. 
These two contributions can  be  distinguished 
in the case of a spin--polarized 
2DEG, where the  MPs are excitations of the spin--$\uparrow$ 
electrons that populate  
the ground state, while   
all  $\sigma = \downarrow$ carriers 
are induced by the 
right--circularly polarized optical pulses. 
In this case, the $\sigma = \uparrow$ contribution 
to Eq. (\ref{ro}) 
describes the MP interaction effects,  
while the 
$\sigma = \downarrow$ term 
describes the X--X interactions. 
The results of \cite{rossi,axt96} 
for the polarization equation of motion are reproduced 
if we add to the Hamiltonian $H$ the electron--phonon interaction
and use Eq. (\ref{Yop}).
The difference here is that both 
the X--X interactions and the interactions between the 
photoexcited carriers and the collective 2DEG excitations 
are described by the same electronic Hamiltonian Eq. (\ref{Hint}).

As demonstrated by Eq. (\ref{Yint}),   
the state $\hat{Y}^\dag | G \rangle $ 
is  a linear combination of 
\{1-MP + 1-LL0-e + 1-LL1-h\}
and \{1-MP + 1-LL1-e + 1-LL0-h\}
 four--particle excitations, into which  
both the LL0 and the LL1 excitons can scatter by interacting with the 
2DEG. 
In the case of  $X_1$, 
the LL1  electron can scatter to 
LL0 by emitting a 
LL0 $\to$ LL1 MP.
Since this  MP 
energy is  close to the 
$e$--LL0 $\to$ $e$--LL1 energy spacing,  
the above scattering  process is almost resonant. 
It therefore 
 provides an efficient 
decay channel of the LL1 exciton 
to a 
\{1-MP + 1-LL0-e + 1-LL1-h\} four--particle excitation
of the ground state 
$| G \rangle$.
All other allowed scattering 
processes are non-resonant. 
The $X_1$  hole can scatter to 
LL0 by emitting a MP, which leads to
a \{1-MP + 1-LL1-e + 1-LL0-h\} four--particle excitation.
The latter state however has energy that is
significantly higher, by an amount 
of the order of $\sim \Omega_c^c+ 
\Omega_c^v$,  from that of 
the initial $X_1$ state.
In the case of $X_0$, 
the LL0 electron can scatter to LL1
by emitting a MP, 
so that 
$X_0 \rightarrow$ \{1-MP + 1-LL1-e + 1-LL0-h\},
or the LL1 hole can scatter to LL0, 
in which case $X_0 \rightarrow$ 
\{1-MP + 1-LL0-e + 1-LL1-h\}.
$|Y_0\rangle$ is thus a linear combination 
of the same final states as $|Y_1\rangle$, also seen from Eq. (\ref{Ysymm}).
However, in this case the energy of all final states 
is   significantly higher than that of the  
initial state $|X_0\rangle$. Therefore, 
the decay of the LL0 exciton 
is suppressed as compared to that of 
the LL1 (or higher) exciton.
As discussed below, this difference in the dephasing of the 
two LL excitons already plays an important role in the linear 
absorption spectra, and has even more profound effects 
on the non-linear optical spectra.

\section{Ultrafast Non-linear Optical Response} 

\label{nlo} 

In this section we obtain the equations of motion 
for the interband polarizations and the intraband 
populations and coherences that determine
the  non-linear optical 
response and identify the contributions 
due to the many--body interactions.  

Within the dipole approximation, the optical spectra are
determined by the polarization of the photo--excited system,
\begin{equation} \label{Ptot}
P(t) = \frac{\mu}{\sqrt{N}} \sum_i P_i(t) \ , \ 
P_i = 
\langle  {\hat X}_i  \rangle,
\end{equation}
which may be obtained from the equations of motion for 
the $X_i$ polarizations $P_i$.
The time evolution of  any operator 
$\hat{O}$ 
is determined by the 
full many--body 
Hamiltonian $H_{tot}(t)$,
which here includes the 2DEG  degrees of freedom that are  
responsible for the dephasing:
\begin{equation} 
\label{eom} 
i \partial_t \langle \hat{O} \rangle 
=\langle [\hat{O},H] \rangle 
- d(t)  \sum_{j} \langle [\hat{O},\hat{X}_{j}^\dag] \rangle 
- d^*(t)  \sum_{j} \langle [\hat{O},\hat{X}_{j}] \rangle. 
\end{equation} 
where $d(t) = \mu E(t) \sqrt{N}$ is the Rabi energy.
Substituting $\hat{O} = \hat{X}_i$ 
in the above equation and using the property 
$[ \hat{X}_i, \hat{X}_{i'}]=0$ 
and Eq. (\ref{Yop}) for the commutator $[\hat{X}_i,H]$ 
we obtain 
the familiar polarization equation of motion: 
\begin{equation}
i \partial_t P_i(t) - \Omega_i P_i(t) +
( 1 - \nu_i) 
\sum_{i' \ne
i} V_{ii'} P_{i'}(t) 
= 
- d(t)
[1 - \nu_i - \sum_{i'} 
\Delta \nu_{ii'} ]
+ \langle \hat{Y}_i \rangle,\label{eom-Pi} 
\end{equation} 
where $\Delta \nu_{ii'} = \langle \Delta \hat{\nu}_{ii'} \rangle$.  
The lhs of the above equation describes the static 
exciton energies and Coulomb--induced LL couplings. 
Two sources of non-linearity are described by 
the terms on the rhs and  
are the subject of this paper. 
The first term describes the
Pauli blocking effects (PSF), which are determined by the 
ground state 
spin--$\downarrow$ electron populations,  
with filling factor $\nu_i$, 
and by the photocarrier
populations and 
 intra--band Raman coherences 
in the  spin--$\downarrow$ electron system, described by 
 $\langle \Delta \hat{\nu}_{ii'} \rangle$, 
Eqs. (\ref{commut-XX-gen}) and (\ref{Dnu}). 
The second term on the rhs of Eq. (\ref{eom-Pi}), 
$\langle \hat{Y}_i \rangle$, 
comes from the interactions between 
$X_i$ and the rest of the carriers in the system: 
X--X, X--MP, and X--2DEG interactions.

The equation of motion for $\Delta \nu_{ii'}$
may be obtained by substituting 
$\hat{O}= [{\hat X}_i, {\hat X}_{i'}^{\dag}]$ in 
Eq. (\ref{eom}) and recalling 
Eq.(\ref{commut-XX-gen}) that expresses
 $\Delta \hat{\nu}_{ii'}$ in terms of the 
exciton operators.  
Calculating the commutator 
$[ H , \Delta \hat{\nu}_{ii'}]$ 
by using the  property 
\begin{equation}
\label{ABC}  
[A,[B,C]] + [C,[A,B]] 
+ [B,[C,A]]=0, 
\end{equation} 
which holds for any operators $A, B, C$,
and 
Eq. (\ref{Yop}) for the commutators $[ \hat{X}_i, H]$,  
we obtain the equation of motion 
\begin{align} \label{eom-Dnuii'} 
i \partial_t \Delta \nu_{ii'} =&
\left(\Omega_i - \Omega_{i'} 
-\frac{i}{T_1}
\right) 
\Delta \nu_{ii'} 
+ \langle [ \hat{X}_i, \hat{Y}^\dag_{i'}] \rangle -
\langle [ \hat{Y}_i, \hat{X}^\dag_{i'}] \rangle\nonumber\\
&+ ( 1 - \nu_{i'}) \sum_{j \ne i'} 
V_{ji'} \Delta \nu_{ij} 
-  ( 1 - \nu_{i}) \sum_{j \ne i} 
V_{ij} \Delta \nu_{ji'} \nonumber\\  
&- d(t) \sum_j \langle [ \Delta \hat{\nu}_{ii'}, 
\hat{X}^\dag_j ] \rangle 
- d^*(t) \sum_j \langle [ \Delta \hat{\nu}_{ii'}, 
\hat{X}_j ] \rangle
\end{align}
where $T_1$ is the  relaxation time.
In the ideal system, 
$\Delta \nu_{ii'} = \Delta \nu_i \delta_{ii'}$
where $\Delta \nu_i$, Eq. (\ref{Dnu}),
is the photo--induced change in the LL filling factor.

A simple model can be derived if we 
restrict to the first two LL states. Using 
Eqs. (\ref{Ysymm}),
(\ref{Yop}), (\ref{ABC}), and   
 Eq. (\ref{commut-XX}) 
for  $\Delta \hat{\nu}_n$ 
we first obtain  the commutator  
\begin{equation} 
\label{commut-nuH}
\left[ \Delta \hat{\nu}_n, H \right] = 
(\delta_{n1} - \delta_{n0}) 
\left( \hat{M}^\dag_n - \hat{M}_n \right), 
\end{equation} 
where we introduced the operator 
\begin{equation}
\label{M-def}  
\hat{M}_n 
= [ \hat{Y},\hat{X}^\dag_n],
\end{equation}
discussed  below.  
Substituting the above 
into Eq. (\ref{eom}) 
and using the commutator 
\begin{equation} 
[\Delta \hat{\nu}_n, \hat{X}^\dag_m ] 
=  \frac{2}{N} \delta_{nm}  \hat{X}^\dag_n,
\end{equation}
obtained by using the second quantization expressions 
for $\Delta \hat{\nu}_n$ and $\hat{X}_m$, 
we obtain the 
equation of motion of 
the photo-induced change in the LL$n$ filling factor, 
 $n=0,1$:  
\begin{equation} 
i \partial_t \Delta \nu_n = 
 \frac{2}{N} \left[ d^*(t) P_n - d(t) P_n^* \right] 
+ ( \delta_{n,1} - \delta_{n,0}) \left[ \langle \hat{M}_n \rangle^{*} 
- \langle \hat{M}_n \rangle \right] - \frac{i}{T_1} \Delta \nu_n 
\label{eom-den} 
\end{equation} 
We note here that the above equation is consistent 
with the  conservation of  the total number of 
photoexcited carriers,
$\Delta \nu_0 + \Delta \nu_1$. 
To see this, we first 
note the following property 
in the case of two LLs 
\begin{equation}
\label{M-symm} 
\hat{M}_n + \hat{M}^{\dag} _{n'} 
= V_{n'n} \left[ ( 1 - \nu_0) \Delta \hat{\nu}_1-
( 1 - \nu_1) \Delta \hat{\nu}_0 \right],
\end{equation} 
where $n' \ne n$ and $n,n'=0,1$, 
which can be obtained after some algebra 
by substituting in the definition 
Eq. (\ref{M-def}) 
of $\hat{M}_n$ the operator $\hat{Y}=\hat{Y}_1 = - \hat{Y}_0$, 
 Eq. (\ref{Yop}),
and using 
the commutator relation
Eq. (\ref{ABC}).
Noting that $V_{10}=V_{01}$, we obtain from the above equation  
that 
$\hat{M}_1^\dag - \hat{M}_1 =
\hat{M}_0^\dag - \hat{M}_0$
and 
\begin{equation} 
i \partial_t \left(\Delta \nu_1 + \Delta \nu_0 \right)= 
- \frac{i}{T_1} \left(\Delta \nu_1 + \Delta \nu_0 \right)  
+ \frac{2}{N} \left[ d^*(t) (P_0 + P_1) - d(t) (P_0^* + P_1^*)  
\right]. 
\end{equation} 
The above
equation demonstrates the conservation 
of the total photoexcited carrier density
when restricting to photoexcitation of the first two LLs.  
The relaxation time $T_1$ is due to the scattering
of the LL0 and LL1 carriers  
to higher LLs as well as their radiative recombination, 
while the last term on the rhs describes the photoexcitation process. 

The 
intraband density matrices 
$\langle [ \hat{Y} , \hat{X}_n^\dag] \rangle$
that enter on the rhs 
of Eq.(\ref{eom-den})  
describe a redistribution of the photoexcited carrier populations 
between the two LLs that is assisted by the MP and the interactions. 
Analogous phonon--assisted effects in the case of undoped semiconductors 
are discussed e.g. in \cite{rossi} and \cite{axt96}.
The corresponding physical processes 
become clear  by calculating 
the above commutator 
using Eqs. (\ref{Yint}) and (\ref{Xq}). 
Restricting for simplicity 
to the first two LLs we obtain that 
\begin{align}
&[\hat{Y}_{int},
\hat{X}_n^\dag] 
 = 
\frac{1}{2 \pi l^2 \sqrt{N}}
\sum_{{\bf q}} \upsilon_q 
\hat{\rho}_{{\bf q}} 
\left[ \phi_{10}({\bf q}) 
\langle G | \hat{X}_{{\bf q}01} \hat{X}^\dag_n  | G \rangle 
\! -\!   \phi_{01}({\bf q}) 
\langle G|  \hat{X}_{{\bf q}10} \hat{X}^\dag_n | G \rangle \right] 
\nonumber \\
&\! + \! (\delta_{n1}\! -\! \delta_{n0}) 
\sum_{{\bf q}} 
\frac{v_{q}}{L^2} 
 \left[  \phi_{10}({\bf q}) 
\hat{X}_{-{\bf q}01}\! -\! 
\phi_{ 01 }({\bf q}) 
\hat{X}_{ - {\bf q}10}\right]^\dag  
\! \left[ 
\phi_{10}({\bf q}) 
\hat{X}_{- {\bf q}01}\! -\! 
\phi_{01}({\bf q}) 
\hat{X}_{ - {\bf q}10}\right]  \nonumber\\ 
&\! + \!  \sum_{{\bf q}} 
\frac{v_{q} \hat{\rho}_{-{\bf q}} }{L^2} 
\left[\delta_{n0} \left( \phi_{01}({\bf q}) 
\hat{\rho}^e_{{\bf q} 01 \downarrow} 
\! -\!  \phi_{10}({\bf q}) \hat{\rho}^h_{{\bf q} 01 \downarrow} 
\right)  \! -\!  \delta_{n1} 
\left( \phi_{10}({\bf q}) \hat{\rho}^e_{{\bf q} 10 \downarrow} 
\! -\!  \phi_{01}({\bf q}) \hat{\rho}^h_{{\bf q}10\downarrow} 
\right) \right]\label{M-2q} 
\end{align} 
The first term on the rhs 
describes the photoexcitation of coherent MPs
and is analogous to the coherent phonon contribution 
in undoped semiconductors
\cite{rossi}. 
Similar to the latter case, it vanishes in the ideal system, 
but is known to contribute in the realistic quantum Hall 
system due to  
disorder, inhomogeneities, and valence band mixing 
\cite{aron-rev,aron-92,raman,marmorkos}. 
Similar to the undoped system \cite{axt96}, the second term describes 
a contribution due to X populations and inter--LL 
exciton coherences, described by the 
density matrices 
$\langle \hat{X}_{{\bf q} nm}^\dag 
\hat{X}_{{\bf q} n'm'} \rangle$. 
The latter exciton 
coherences and populations come from the photoexcited carriers, 
while in the 2DEG system there are additional coherences
due to the ground state 2DEG excitations,   
described by the last term. In addition to 
 interactions 
among the photoexcited carriers 
similar to the undoped system \cite{rossi}, 
the latter term 
describes the scattering and correlations 
between MPs and spin--$\downarrow$ carriers, 
which lead to the relaxation of the photoexcited carriers 
due to MP emission and absorption.
 
As can be seen from Eq. (\ref{M-2q}),  
the intraband scattering processes are
described by density matrices of the form 
$\langle \hat{X}^\dag \hat{X} \rangle,
 \langle \hat{\rho}^e_{\sigma} 
\hat{\rho}^e_{\downarrow} \rangle, 
\langle \hat{\rho}^e_{\sigma} 
\hat{\rho}^h_{\downarrow} \rangle,$
and $ 
\langle \hat{\rho}^h_{\downarrow} 
\hat{\rho}^h_{\downarrow} \rangle$.
In the case of the spin--$\uparrow$ polarized 2DEG, 
the MP contributions are described by the 
density matrices 
$ \langle \hat{\rho}^e_{\uparrow} 
\hat{\rho}^e_{\downarrow} \rangle$ and  
$\langle \hat{\rho}^e_{\uparrow} 
\hat{\rho}^h_{\downarrow} \rangle$, 
which vanish in the undoped system in the case of 
right--circularly polarized light and 
are analogous to the phonon--assisted 
intraband density matrices \cite{rossi}.  
They relate an initial state consisting of a photoexcited electron 
or hole to a final state consisting of an electron or hole 
plus a MP and describe the effects of carrier scattering 
by MP emission or absorption. 

The number of independent density 
matrices can be reduced by noting that 
the e--h pair creation operators may be expressed in terms of exciton 
operators: 
\begin{equation} 
\hat{e}^\dag_{k n \downarrow} 
\hat{h}^\dag_{-k' m \downarrow} 
= \frac{1}{\sqrt{N}} \sum_{{\bf q}} 
\hat{X}^\dag_{{\bf q} n m} e^{- i q_x (k + k') l^2/2} 
\delta_{q_y,k-k'}, 
\label{eh-X} 
\end{equation} 
obtained from Eq. (\ref{Xq}).
Similar to the DCTS, 
further reductions can be obtained by noting that, as discussed below, 
only many--body states with one valence band  hole
contribute to the above intraband density 
matrices in the case of the third--order
non-linear optical response, 
and thus the density matrix
$\langle \hat{h}^\dag \hat{h}^\dag 
\hat{h} \hat{h} \rangle$ 
can be neglected to this order. 
Furthermore, 
in the case of the spin--$\uparrow$ polarized 2DEG 
excited with right--circularly polarized light,  
spin--$\downarrow$ 
carriers are only created via the photexcitation,  
and thus the density matrix
$\langle \hat{e}^\dag_{\downarrow} \hat{e}^\dag_{\downarrow}  
\hat{e}_{\downarrow} \hat{e}_{\downarrow} \rangle$ 
contributes to higher order, 
similar to the undoped system. 
In this case, only
states with one spin--$\downarrow$ 
electron or hole contribute,  
and  we 
obtain 
by using 
Eq. (\ref{eh-X})
and denoting by $n_{e \downarrow}$ and 
$n_{h \downarrow}$  the number operators
of the spin--$\downarrow$ carriers that 
\begin{equation} 
\hat{\rho}^e_{{\bf q}nn'\downarrow}
= \hat{\rho}^e_{{\bf q}nn'\downarrow} 
n_{h \downarrow}
= \frac{1}{\sqrt{N}} 
\sum_{{\bf q}'m} e^{i ({\bf q} \times {\bf q}')_z l^2/2}
\hat{X}^\dag_{{\bf q}'nm} 
\hat{X}_{{\bf q}' - {\bf q}n'm} 
\end{equation} 
and 
\begin{equation} 
\hat{\rho}^h_{{\bf q}nn'\downarrow} 
= \hat{\rho}^h_{{\bf q}nn'\downarrow} 
n_{e \downarrow} = 
 \frac{1}{\sqrt{N}} 
\sum_{{\bf q}'m} e^{-i ({\bf q} \times {\bf q}')_z l^2/2}
\hat{X}^\dag_{{\bf q}'mn} 
\hat{X}_{{\bf q}' - {\bf q}mn'}. 
\end{equation} 
The above expressions can be used 
to show that, 
in the case of spin--polarized 2DEG
(e.g. for filling factors $\nu = 1/m$ ($m=$ integer)
or integer $\nu$)    
and right--circular polarization, 
the independent intraband density 
matrices have the form 
$\langle \hat{X}^\dag \hat{X} \rangle$ 
and $\langle \hat{\rho}^e_{\uparrow} \hat{X}^\dag \hat{X} \rangle$.
More details will be presented elsewhere.  
For other filling factors, 
all density matrices that enter in Eq. (\ref{M-2q}) 
must be calculated to obtain the third--order 
response, with the exception of 
$\langle \hat{h}^\dag \hat{h}^\dag \hat{h} \hat{h} \rangle$
which contributes to higher order.

To conclude this section, we note from the above equations 
of motion that 
the effects of the interactions on the non-linear optical response 
are described by the interband density matrix $\langle \hat{Y} \rangle$ 
and the intraband density matrix $\langle \hat{M}_n \rangle$.
Due to the many--body nature of this strongly correlated 
 system, approximations are needed 
in order to calculate these interaction--induced effects. 
In the undoped system, 
the DCTS cumulant expansions 
separate  the coherent from the 
incoherent and the correlated from the uncorrelated contributions. 
In the case of the 2DEG, 
we introduced in \cite{kara} an analogous decoupling scheme 
based on a  decomposition of the many--body 
semiconductor state $| \psi\rangle$
that 
evolves from the ground state $| G \rangle$ according to the
Schr\"{o}dinger equation for 
the Hamiltonian $H_{tot}(t)$.
This method is summarized in the following section  
and then used to 
derive a factorization scheme for the 
density matrix equations of motion. 

\section{Interaction Effects} 
\label{correl} 

In this section we discuss a decomposition of the 
photoexcited many--body wavefunction into correlated and 
uncorrelated contributions,
which we use in the next section 
to devise approximations 
for  treating the interaction--induced density matrices 
$\langle \hat{Y}_n \rangle$ and 
$\langle [\hat{Y}_n, \hat{X}^\dag_m] \rangle$.
Similar to the DCTS, 
we expand in terms of the  
optical field and calculate the third--order polarization, 
which is expected to describe the 
non-linear optical  signal 
when
the photoexcited carrier density is
small and the X--cold 2DEG  correlations  prevail. 

As in the theoretical approaches of \cite{axt96,ost-98},
we note the one to one correspondence
between the photon absorption/emission processes and the $e$--$h$
pair creation/destruction. However, since 
here a 2DEG is present
prior to the photoexcitation, when  following the effects of the applied
fields we count the number of valence band
holes in a given state.
Therefore, we use the 
shorthand notation 0--$h$, 1--$h$, 2--$h$ ... to label the   states,
and it is clear that states with three or more holes do not
contribute to the third--order non-linear polarization
\cite{mukamel-book}. We can then decompose the optically--excited
state $|\psi\rangle$ according to
\begin{equation} \label{psiexpansion}
| \psi \rangle = | \psi_0 \rangle + | \psi_1 \rangle + | \psi_2
\rangle,
\end{equation}
where $| \psi_i \rangle$, $i=0,1,2$, describes the contribution of
the $i$--$h$ states, 
with initial condition $| \psi_i(-\infty) \rangle = \delta_{i,0} |
G \rangle$. From now on, we refer to operators that change the 
number of holes as interband and to operators that leave the number of
holes unchanged as intraband. 
 
\subsection{Linear response} 
\label{linear} 

To lowest order in the optical field, only the 
1--$h$ state
$ | \psi_1 \rangle$ is photoexcited. 
We separate out the magnetoexciton contribution 
to this state 
by introducing the decomposition 
\begin{equation}
\label{1hole} | \psi_{1L} \rangle = \sum_i \frac{P_{i}^{L}}{1 - \nu_i}
| X_i \rangle +
| \bar{\psi}_{1L} \rangle,
\end{equation}
where  $|\bar{\psi}_{1L}\rangle$ is the 
\{1-$h$/2DEG$^\ast$\} contribution, defined by the condition 
$\langle X_i | \bar{\psi}_1\rangle$ = 0, 
that describes  the incoherent contributions 
due to the X--2DEG interactions. 
From now on, 2DEG$^\ast$ denotes excited 2DEG configurations.
The exciton amplitude
\begin{equation}
\label{pol} P_i^{L} =  \langle X_i  | \psi_{1L} \rangle 
\end{equation}
coincides with the linear 
polarization, 
whose equation of motion is obtained
by linearizing Eq. (\ref{eom-Pi}).
As shown in \cite{kara}, 
we obtain the equation of 
motion for $|\bar{\psi}_{1L} \rangle$
by substituting the decomposition Eq.\ (\ref{1hole}) into the
linearized Schr\"odinger equation 
for $| \psi \rangle$  
and 
using the equation of motion for the linear polarization
$P_i^L$: 
\begin{equation} 
\label{bar1hole}
i \partial_t |\bar{\psi}_{1L} \rangle = H |\bar{\psi}_{1L} \rangle 
+ \sum_{i} \frac{1}{1 - \nu_i} 
\left[ P_i^L |Y_i \rangle
- \bar{P}_i^L | X_i \rangle \right]
\end{equation}
where the amplitude 
\begin{equation}
 \bar{P}_i^L 
= \langle Y_i | \bar{\psi}_{1L}
\rangle = \langle Y_i | \psi_{1L}
\rangle 
\end{equation} 
coincides with $\langle \hat{Y}_i \rangle$ to first order in the optical 
field.
If we restrict to the LL0 and LL1 states, 
Eq. (\ref{Ysymm}) applies and 
we have that
\begin{equation} 
\bar{P}_1^L 
= - \bar{P}_0^L 
= \bar{P}^L. 
\end{equation} 
The amplitude
$\bar{P}_i^L$ describes the time evolution 
of the X+MP states that 
contribute to $| Y \rangle$ 
and corresponds to a X+MP coherence, analogous to the 
X+phonon coherence in undoped semiconductors \cite{rossi,axt96}.

To obtain the equation of motion for $\bar{P}^L$, we must consider the 
action of the Hamiltonian on the state $| Y \rangle$ introduced in section 
\ref{MP-MEX}. Similar to Eq.\ (\ref{HonX}) that defines the latter state,
we introduce a new state
$|Z\rangle$ 
orthogonal to all the X states as well as to $| Y \rangle$ 
as follows \cite{kara}: 
\begin{equation}
\label{HonY} 
H |Y \rangle = \bar{\Omega} |Y \rangle +   
W \left( \frac{| X_1 \rangle}{1 - \nu_1} 
- \frac{| X_0 \rangle}{1 - \nu_0} 
\right) 
 + |Z \rangle , 
\end{equation}
where we obtain after some algebra \cite{kara} 
from the above orthogonality requirements and  
after using Eqs. (\ref{X-ortho}), (\ref{HonX}),
and (\ref{Ysymm}) 
\begin{equation}
\label{Yen} \bar{\Omega} =\frac{\langle Y|  H | Y \rangle}{
\langle Y|Y \rangle} \ , \ 
W = \langle Y | Y \rangle.
\end{equation} 
By projecting the state $\langle Y|$ to Eq. (\ref{bar1hole}) 
and using Eqs. (\ref{HonY}), (\ref{Ysymm}), and the orthogonality 
among the above states,   
 we obtain the 
equation of motion 
for $\bar{P}^L$: 
\begin{equation} 
i \partial_t \bar{P}^L 
= ( \bar{\Omega} - i \gamma) 
\bar{P}^L + 
W \left( \frac{ P_1^L}{1 - \nu_1} 
- \frac{P_0^L}{1 - \nu_0}
\right)
+ Z^{L}
\end{equation} 
where 
$  Z^{L} = \langle Z | \bar{\psi}_{1L} \rangle$ 
and $\gamma$ is the dephasing rate. 

By continuing the above orthogonalization 
procedure,
we create 
a Lanczos  basis \cite{schaf-book,haydock}  
of strongly  
correlated orthogonal states $ | Z^{(n)}\rangle$, where 
$ | Z^{(0)}\rangle = | Y \rangle$ and  
$| Z^{(1)}\rangle = | Z \rangle$, 
from the recursive relation 
obtained 
by acting with the Hamiltonian $H$ on the previous state, and then
orthogonalizing the result with respect to all the existing basis
states \cite{kara,haydock}: 
\begin{equation} 
H | Z^{(n)} \rangle = 
\bar{\Omega}^{(n)} 
| Z^{(n)} \rangle + 
W^{(n)} 
| Z^{(n-1)} \rangle + 
 | Z^{(n+1)} \rangle
\end{equation}
where 
\begin{equation} 
\bar{\Omega}^{(n)} 
= 
\frac{\langle  Z_{n}| H | Z_{n} \rangle}{\langle  Z_{n}| Z_{n} \rangle}
\ , \ 
W^{(n)} =
\frac{\langle  Z_{n}| Z_{n} \rangle}{\langle  Z_{n-1}| Z_{n-1} \rangle}.
\end{equation} 
By projecting the state $\langle  Z_{n}|$ to Eq. (\ref{bar1hole})
we obtain the equations of motion 
for the coherences 
$Z^{(n)}_L = \langle  Z_{n}|
\bar{\psi}_{1L} \rangle$: 
\begin{equation} 
i \partial_t  Z^{(n)}_L = 
\left( \bar{\Omega}^{(n)}
- i \gamma_{n} \right)  Z^{(n)}_L 
+ W^{(n)} 
Z^{(n-1)}_L + Z^{(n+1)}_L. 
\end{equation} 
As discussed in \cite{kara}, 
by taking the Fourier transform of  
the above system of equations, we can obtain a continued fraction
expansion of the polarization,
which describes a non--Markovian dephasing.
This approach is analogous to the numerical calculations of the 
2DEG dynamical 
structure factor  \cite{QHE1,QHE2}.  
The above hierarchy 
can be truncated when convergence is
reached, which 
becomes more rapid  with increasing damping rates or,     
in the case of an $N$--electron system, 
after performing $N$ iterations.
In the QHE literature, numerical calculations 
of the $N$--electron spectral functions 
have been shown to 
extrapolate to   the 
$N \rightarrow \infty$ limit
for a relatively small $N$ \cite{QHE1,apalkov}. 
Compared to an expansion in terms of 
noninteracting X+MP states, 
as in Eq. (\ref{Yint}), 
the correlated Lanczos states are advantageous 
when  the different momentum 
contributions are strongly coupled, 
or for obtaining a simple solution such as the average polarization 
model discussed below. 

\begin{figure}[t]
\begin{center}
\includegraphics*[width=7cm]{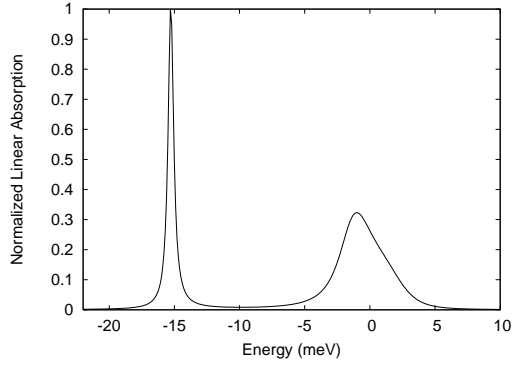}
\caption{\label{LA}
Linear absorption spectrum with parameters 
as in \cite{kara,from-02-prb,dani} 
chosen to fit the experimental results
at $B=7T$
($\sqrt{W}=3meV$, $\Gamma_0=\Gamma_1=0.12meV$, 
$\gamma=3.8meV$, 
$\nu_0=0.34$, $\nu_1=0$).
}
\end{center}
\end{figure}

In Fig.\ref{LA}  we show the linear absorption spectrum obtained 
by retaining the states $|X_n \rangle, n=0,1,$ 
and $| Y \rangle$. The effect of the higher Lanczos states, 
which describe the ``bath'' that leads to exciton dephasing
and do not couple directly to the exciton, 
is taken into account by introducing the dephasing rate $\gamma$ 
of $\bar{P}^L$ that 
describes 
the coupling between the exciton and the ``bath''. 
This dephasing 
is due to  X+MP scattering (see Eq.(\ref{Yint}))
and the emission of other  2DEG excitations. 
Noting the analogy between a MP and an X discussed in
Section (\ref{MP-MEX}),  
one can make an analogy 
between the X--MP scattering 
described by Eq.(\ref{Yint})
and the X--X scattering  in undoped semiconductors. 
As shown in \cite{shah00}, in the case of magnetoexcitons the latter
can be described by  a dephasing rate 
for strong X--X interactions and  
leads to the average polarization model
\cite{chemla-99,schaf-96,shah00}.
%
% FIG.1: LINEAR ABSORPTION SPECTRUM
The main feature in Fig.\ref{LA} is the strong LL1 
broadening and asymmetric lineshape, which is 
due to 
$\bar{P}^L$
and cannot be obtained by introducing a polarization dephasing time. 
Even though  the time evolution of the X+MP states, 
described by $\bar{P}^L$, determines the 
lineshape of the LL1  peak, 
it only plays a small role at the 
LL0 frequency.  
To interpret this difference 
we note that, as can be seen 
by using Eq. (\ref{Yint}) 
to calculate the state $| Y \rangle$, 
the main 
contribution to $\bar{P}^L$ comes from 
\{1-MP + 1-LL0-e + 1-LL1-h\}
four--particle excitations, 
while all other contributions 
are non-resonant with 
the LL0 and LL1 excitons.  
Even though $\bar{P}^L$  couples equally 
to both X amplitudes $P_0^L$ and $P_1^L$, 
it dominates the 
dephasing of $P_1^L$
since the above four--particle states  
 have energy comparable 
to that of $X_1$ in the case of 
an inter--LL MP. 
In contrast, $X_0$ has significantly smaller energy, 
and thus 
the inter--LL MP plays a minor role 
in the broadening of the LL0 exciton peak.

\subsection{Second order processes}

In this section we consider the two--photon non-linear optical processes
that lead to 
the photoexcitation of the 2--$h$ 
state 
$| \psi_2 \rangle$ and the 0--$h$ state 
$ |\psi_0 \rangle.$  
 By separating out the contribution of the  states 
$| X_{i} X_{i'} \rangle = {\hat
X}^\dag_i {\hat X}^\dag_{i'} | G \rangle$, 
which  describe a pair of non-interacting
Xs, and $\hat{X}_i^\dag | \bar{\psi}_{1L} \rangle$,  
which describes  a noninteracting 
pair of X and X+MP 1--$h$ states, 
we arrive at the decomposition to $O(E^4)$
\begin{equation}
 | \psi_2\rangle =  \frac{1}{2} \sum_{ii'} \frac{P_{i}^L
P_{i'}^L}{(1 - \nu_i )(1 - \nu_{i'})}  
 | X_{i} X_{i'} \rangle + 
 \sum_{i} \frac{P_{i}^L}{1 - \nu_i} {\hat X}^{\dag}_{i} |
\bar{\psi}_{1L} \rangle + |\bar{ \psi}_2\rangle  
\label{2hole}
\end{equation}
where $|\bar{ \psi}_2\rangle$ describes the correlated 
X--X and X--X+MP contributions and satisfies the equation of motion 
\cite{kara} 
\begin{equation}
 i \partial_t | \bar{\psi}_2 \rangle - H |
\bar{\psi}_2 \rangle = \frac{1}{2} \sum_{ii'} 
\frac{P_{i}^L P_{i'}^L [{\hat Y}^\dag_i , {\hat
X}^{\dag}_{i'}] | G \rangle
}{
( 1 -\nu_i) (1 -\nu_{i'})}
+ \sum_{i} \frac{1}{1 - \nu_i}\left[ P_{i}^L {\hat
Y}_{i}^{\dag}- \bar{P}_{i}^L {\hat X}^{\dag}_{i}\right]
|\bar{\psi}_1 \rangle.\label{bar2hole}
\end{equation}
We note that, in the above equation of motion, there are no terms 
proportional to $d(t)$ 
and therefore the decomposition Eq. (\ref{2hole})  eliminates 
all contributions to $|\psi_2 \rangle$ 
that are proportional to the excitonic amplitudes 
$P_i^L$ ( whose time derivative is proportional to $d(t)$).  

In addition to the photoexcitation of the above  2--$h$ many--body state,  
the two--photon processes of excitation of a 1--$h$ state   
and then de-excitation 
of an $e$--$h$ pair,  
possibly  accompanied
by the scattering of 2DEG excitations, 
leads to a second order contribution to the 
0--$h$ state $ | \psi_0 \rangle$. 
We split the latter 
state  into the contribution of the ground state $|G\rangle$,
with amplitude $\langle G | \psi \rangle$,
where the 2DEG is not excited during the 
excitation--deexcitation process,  
and the photoexcited \{0-$h$/2DEG$^\ast$\} contribution. 
We further decompose the latter contribution 
into an uncorrelated part $\propto \hat{X}_i | \bar{\psi}_{1L} 
\rangle$, where the deexcited exciton $\hat{X}_i$ 
does not interact with the $| \bar{\psi}_{1L} \rangle$ carriers, 
and a correlated contribution 
$| \bar{\psi}_0 \rangle $:
\begin{equation}
| \psi_0 \rangle = \langle G | \psi \rangle \, | G
\rangle - 
\sum_{i} \frac{P_{i}^{L*}}{1 - \nu_i}
{\hat X}_{i} | \bar{\psi}_{1L} \rangle + |
\bar{\psi}_0 \rangle + O(E^4), 
\label{0hole} 
\end{equation}
where  the  2DEG$^\ast$ state $|\bar{\psi}_0 \rangle$, 
$ \langle G|\bar{\psi}_0
\rangle  = 0$, 
satisfies the equation of motion \cite{kara} 
\begin{multline} \label{bar0hole}
i \partial_t | \bar{\psi}_0 \rangle -
 H | \bar{\psi}_0 \rangle=
\sum_{ii'} \frac{P_{i}^{L*} P_{i'}^L}{
(1 - \nu_i) (1 - \nu_{i'})} \hat{X}_i |Y_{i'} \rangle 
+\sum_{i}\frac{P_{i}^{L*} {\hat Y}_{i} -
{\bar P}^{L*}_{i} {\hat X}_{i}}{1 - \nu_i} 
| \bar{\psi}_1 \rangle \\
-\sum_{ii'} \frac{P_{i}^{L*} \bar{P}^L_{i'}}{
( 1 - \nu_i) ( 1 - \nu_{i'})}  
\hat{X}_i | X_{i'}\rangle
- d^*(t) \ \sum_{ii'} \frac{ P_{i'}^L}{1 - \nu_{i'}}  
\left( [\hat{X}_i, \hat{X}^\dag_{i'}]
 - \delta_{ii'} \right) | G \rangle.
\end{multline}
The first term in Eq.\ (\ref{bar0hole})
describes the  photo--excitation of the 2DEG via 
the second--order interaction--assisted  process 
where the exciton $X_{i'}$, photo--excited 
with amplitude $P_{i'}^L$,
scatters with the 2DEG into the state
$|Y_{i'}\rangle$, and then the exciton $X_i$ is deexcited 
with amplitude $P_{i}^L$. The above process leaves the system 
in a 2DEG$^*$ 0--h state. It is analogous to the photoexcitation 
of coherent phonons in  undoped semiconductors,
and dominates the 
inelastic light scattering  spectra of the 2DEG. 
\cite{aron-rev,raman}
The second term on the rhs of Eq.\
(\ref{bar0hole}) describes the scattering of $X_i$ with the
carriers in $|\bar{\psi}_1\rangle$
during its de-excitation. The rest of the terms 
describe the possibility to create  2DEG  excitations 
by photoexciting an exciton whose hole 
then recombines with a 2DEG electron (Raman process). 
In the case of the spin--$\uparrow$ polarized 2DEG and 
right--circularly polarized light, 
the latter Raman process vanishes since there are no 
spin--$\downarrow$ electrons in the ground state.

Below we use the above decompositions of the photoexcited 
many--body wavefunction 
$| \psi \rangle$ 
in order to describe the  interaction--induced contributions
to Eqs. (\ref{eom-Pi}) and (\ref{eom-den}). 
 These  decompositions provide a 
way of separating out the uncorrelated from the correlated parts
in a general correlated system, where a Hartree--Fock noninteracting state 
may not be an appropriate reference state as in
undoped semiconductors.  
This  separation also motivates a factorization 
of the corresponding density matrices 
that applies not only to undoped semiconductors,
but also to systems with  strongly correlated 
ground states, where Wick's theorem does not apply.
Similar to the DCTS,
 our method provides a systematic way of 
 identifying the parts that can be factorized 
and the new  intraband dynamical variables 
that cannot be expressed in terms 
of interband coherences due to the incoherent processes. 
Importantly, it allows us to treat 
both coherent and incoherent processes 
in strongly correlated systems, where the coupling 
between the photoexcited carriers and the ``bath'' (here the 2DEG) 
leads to new dynamics governed by slow/low energy 
``bath'' collective excitations.
Finally, our method raises the possibility 
of devising  new  approximations, 
obtained by projecting the many--body wavefunctions
in  an appropriate basis 
of strongly correlated states or operators, and then using this expansion 
to evaluate the correlated contributions to the density matrices.   

\subsection{Intraband density matrix} 

In this section 
 we turn to the equation of motion for the density matrix  $\langle 
\hat{M} \rangle$, where 
 $\hat{M}$ is any intraband  operator that 
does not change the number of holes.
Furthermore, we assume that 
$\langle G | \hat{M} | G \rangle =0$,   
as is the case for the operator
$[\hat{Y}_i , \hat{X}^\dag_j]$ discussed above.  
Substituting the above 
decomposition of the many--body state 
$| \psi \rangle$ we obtain for the 
average value 
$\langle \psi | 
\hat{M} | \psi \rangle$ by 
keeping terms up to second order in the optical field 
\begin{multline}\label{Mc-states} 
\langle \hat{M} \rangle = 
\langle \hat{M} \rangle_c
+ \sum_{ij} \frac{P_i^{L*} P_j^L}{(1 - \nu_i) (1 - \nu_j)} 
\langle X_i | \hat{M} | X_j \rangle \\
+ \sum_i \frac{P_i^{L*}}{1 - \nu_i}
\langle G | [ \hat{X}_i, \hat{M}] | \bar{\psi}_{1L} \rangle 
+ \sum_i \frac{P_i^{L}}{1 - \nu_i}
\langle \bar{\psi}_{1L}  | [ \hat{M}, \hat{X}_i^\dag] | G \rangle 
+ O(E^4), 
\end{multline} 
where $\langle \hat{M} \rangle_c$ is the correlated contribution, given by 
\begin{equation} 
\label{Mc-def}
\langle \hat{M} \rangle_c
= \langle G | \hat{M}| \bar{\psi}_0 \rangle 
+\langle \bar{\psi}_0 | \hat{M} | G \rangle 
+ \langle \bar{\psi}_{1L} |  \hat{M}
| \bar{\psi}_{1L} \rangle.
\end{equation} 
This above result corresponds to an intraband  density matrix 
decomposition into a factorizable part and 
a correlated part   $\langle \hat{M} \rangle_c$. 
The second term on the rhs is the coherent contribution, 
which similar to the undoped system can be expressed as a product 
of  exciton polarizations, 
while the rest of the terms describe the 
incoherent contributions and 2DEG 
photoexcitation processes. 
The above decomposition corresponds to a 
projection of the exciton states $| X_i \rangle$. 

We can obtain the equation of motion for  $\langle \hat{M} \rangle_c$
either from Eq. (\ref{Mc-def}) 
by using the equations of motion Eqs. (\ref{bar1hole})
and (\ref{bar0hole}) 
and the orthogonality 
$\langle G | \hat{M} | G \rangle =0$
or from Eqs. (\ref{eom}) 
for $\langle \hat{M} \rangle$
and Eqs. (\ref{eom-Pi}) and 
(\ref{bar1hole}): 
\begin{align} 
i & \partial_t \langle \hat{M} \rangle_{c} 
- \langle [ \hat{M}, H] \rangle_{c}  
+ i \gamma_M \langle \hat{M} \rangle_{c} =  
\sum_{i'j'} \frac{ P_{i'}^{L*} 
P_{j'}^L}{(1 - \nu_{i'}) 
( 1 - \nu_{j'})} 
\nonumber \\ &
\times \left[ i ( \Gamma_{i'} + \Gamma_{j'} - \gamma_M)  \langle X_{i'} 
| \hat{M} | X_{j'} \rangle 
+ \langle  Y_{i'}| \hat{X}^\dag_{j'} 
 \hat{M} | G \rangle -
\langle G | \hat{M} 
 \hat{X}_{i'}  | Y_{j'} \rangle 
\right] 
\nonumber  \\ &
+  \sum_{i'} \frac{ i(\Gamma_{i'} + \gamma - \gamma_M)}{1 - \nu_{i'}}
\left[ P_{i'}^L
\langle \bar{\psi}_{1L} | [ \hat{M}, \hat{X}^\dag_{i'} ] | G \rangle
+ P_{i'}^{L*}
\langle G |  [ \hat{X}_{i'},\hat{M}]  |  \bar{\psi}_{1L} \rangle
\right] 
\nonumber \\ &
+ \sum_{i'j'} \frac{P_{i'}^L \bar{P}_{j'}^{L*}}{
(1 - \nu_{i'})(1 - \nu_{j'}) }
\langle X_{j'} | \hat{X}^\dag_{i'} \hat{M} | G \rangle 
-  \sum_{i'j'} \frac{\bar{P}_{i'}^L P_{j'}^{L*}}{
(1 - \nu_{i'})(1 - \nu_{j'}) }
\langle G |\hat{M} 
 \hat{X}_{j'} | X_{i'} \rangle \nonumber \\ &
+ \sum_{i'} \frac{P_{i'}^L}{1 - \nu_{i'}} \langle \bar{\psi}_{1 L} | 
[ \hat{M}, \hat{Y}^\dag_{i'}] | G \rangle 
+ \sum_{i'} \frac{P_{i'}^{L*}}{1 - \nu_{i'}} \langle G| 
[ \hat{M}, \hat{Y}_{i'}] | \bar{\psi}_{1 L}   \rangle 
\nonumber \\ &
+ \sum_{i'} \frac{\bar{P}_{i'}^{L*}}{1 - \nu_{i'}} 
\langle  G | 
[  \hat{X}_{i'}, 
\hat{M}] | \bar{\psi}_{1 L}  \rangle 
+ \sum_{i'} \frac{\bar{P}_{i'}^{L}}{1 - \nu_{i'}} \langle 
 \bar{\psi}_{1 L} | 
[ \hat{X}^\dag_{i'}, \hat{M}] |G   \rangle 
\nonumber \\ &
+ \sum_{i'j'} \frac{ d(t)P_{j'}^{L*}}{1 - \nu_{j'}}
\langle G | [ \hat{X}_{j'}, \hat{X}^\dag_{i'}] \hat{M} | G \rangle 
-  \sum_{i'j'} \frac{ d^*(t)P_{j'}^{L}}{1 - \nu_{j'}}
\langle G | \hat{M} [ \hat{X}_{i'}, \hat{X}^\dag_{j'}] | G \rangle.  
\label{eom-Mcorr} 
\end{align} 
The above equation of motion applies for any intraband operator 
$\hat{M}$, $\langle G | \hat{M} | G \rangle$ = 0, 
such as the operators that contribute 
to Eq. (\ref{M-2q}),   
and describes contributions due to the possible photoexcitation 
of a 2DEG coherence associated with the 2DEG$^\ast$ state 
$\hat{M} | G \rangle$ 
 as well as due to deviations 
from the factorization  
Eq. (\ref{Mc-states})
induced by incoherent processes 
that involve the photoexcited carriers.
The above equation of motion can be evaluated e.g. by expanding 
the operator $\hat{M}$ in  the strongly correlated 
basis 
$  Z^{(n)} \rangle \langle Z^{(m)}| $,  
 $ | X_{n} \rangle \langle Z^{(m)}| $,  and 
 $ | X_{n} \rangle \langle X_{m}| $
and a similar basis of
0--h 2DEG states discussed below.

\subsection{Interband density matrix} 

We now consider 
the equation of motion for the density matrix
$\langle 
\hat{Y} \rangle$, where $\hat{Y}$ 
is any interband operator that 
creates an $e$--$h$ pair with the simultaneous  scattering 
of any number of other electrons or holes 
(e.g. the operator $\hat{Y}_{int}$, Eq. (\ref{Yint})). 
By using the above decomposition of the photoexcited  many--body
state $| \psi \rangle$, 
we obtain after some algebra 
by keeping terms up to third order in the optical 
field
the following 
decomposition of 
the interband density matrix $ \langle \hat{Y} \rangle$ 
into correlated and uncorrelated contributions \cite{kara}:
\begin{align}\label{interactions} 
\langle {\hat Y} \rangle =& 
\sum_i \frac{P_{i}^{L*}}{ 1 - \nu_i} \langle G | [ \hat{X}_i , \hat{Y}] 
| \psi_2 \rangle 
+ \sum_{i} \frac{P_{i}^{L}}{1 - \nu_{i}}  
\langle [ \hat{Y},\hat{X}^\dag_{i}] \rangle_c  
\nonumber\\& 
+ \frac{1}{2} \,  \sum_{ii'} 
\frac{
P_{i}^L \, P_{i'}^L}{
( 1 - \nu_{i}) ( 1 - \nu_{i'})}   
\langle \bar{\psi}_{1L} | [ [ \hat{Y}, \hat{X}_i^\dag],  {\hat
X}^{\dag}_{i'}] | G \rangle 
+  \langle \hat{Y} \rangle_c 
\end{align}
where 
\begin{equation}\label{Y-c}
 \langle \hat{Y} \rangle_c 
 = \langle \psi | G \rangle \, \bar{P}^{L} 
+ \langle Y | \bar{\psi}_{1 NL} \rangle
+ \langle {\bar \psi}_0 | {\hat Y} | {\bar \psi}_{1L} \rangle+ 
\langle {\bar \psi}_{1L} | {\hat Y} | {\bar \psi}_2 \rangle,
\end{equation}
and we introduced the non-linear 1--$h$ state 
\begin{equation} 
| \bar{\psi}_{1 NL}\rangle
= | \psi_{1} \rangle - 
| \psi_{1 L} \rangle + \sum_{i} \frac{P_{i}^{L *}}{1 - \nu_{i}}
\hat{X}_{i} | \psi_2 \rangle
- \sum_{i} \frac{P_{i}^{L}}{1 - \nu_{i}} 
\hat{X}_{i}^\dag | \bar{\psi}_{0} \rangle.
\end{equation} 
The first term on the rhs of Eq. (\ref{interactions}) 
describes the coherent X--X interaction and correlation effects
analogous to the undoped system, discussed in the following section. 
The second term 
describes the contribution due to 
 scattering 
of the optical  polarization with 
the correlated intraband contributions, 
i.e. the intraband coherences and 
incoherent populations  discussed above. 
The above two terms treat the effects of coherent X--X interactions 
and the intraband excitation and incoherent population processes. 
The third term in Eq. (\ref{interactions}) 
describes X--X interactions 
accompanied by the shake--up of 2DEG excitations, 
while the last term describes the correlated contribution
to the non-linear polarization.
As can be seen from Eq. (\ref{Y-c}), 
the dynamics of the first two terms of the latter contribution 
is governed by the dephasing of the 
X+MP state $| Y \rangle$
 while the last two non-linear terms 
are determined by the 
\{1-$h$/2DEG$^\ast$\} state 
$|\bar{\psi}_{1L} \rangle$
and describe incoherent correlated interband  contributions. 
The equation of motion for $\langle \hat{Y} \rangle_c$ 
can be obtained as above and describes the X dephasing.

\section{Coherent X--X correlations} 
\label{XX} 

In this section we  make the connection between the above result 
for the coherent X--X interaction contribution to the non-linear 
polarization (first term 
on the rhs of Eq. (\ref{interactions})), 
described by the amplitude of the 2--$h$ state 
$\langle G| [\hat{X}_i, \hat{Y}] | \psi_2 \rangle
= \langle [\hat{X}_i, \hat{Y}] \rangle + O(E^5),$
and the familiar expressions that describe X--X 
correlations in undoped semiconductors \cite{axt98,axt96}. 
Using Eq. (\ref{Yint}) and restricting to the 
first two LLs  
we obtain for $\hat{B} = [\hat{X}_1,\hat{Y}]$
\begin{equation} \label{B}
\hat{B}
 =  \sum_{{\bf q} } \frac{\upsilon_q}{L^2} 
\left[ \phi_{01}(- {\bf q}) \hat{X}_{-{\bf q} 01}
\!-\! \phi_{10}(- {\bf q}) \hat{X}_{-{\bf q} 10} \right] 
\left[\phi_{01}( {\bf q}) 
\hat{X}_{{\bf q} 01}  \!-\! \phi_{10}({\bf q}) \hat{X}_{{\bf q} 10} \right]
\end{equation} 
and $[\hat{X}_0,\hat{Y}]=- \hat{B}$.
As can be seen from the above equations, 
the coherent X--X interactions are determined by 
 2--X density matrices of the form 
$\langle \hat{X}_i \hat{X}_j \rangle$, similar to the undoped system, 
where $i,j$ describe excitons with finite total momentum. 
Using the decomposition Eq. (\ref{2hole}) 
we obtain 
for any 2--h state $| B \rangle$ 
(e.g. the states $ [ \hat{Y}^\dag, \hat{X}_i^\dag] | G \rangle$ 
that determine the coherent X--X contribution
to the non-linear polarization)  
\begin{equation} 
\label{2X} 
\langle B |\psi_2
\rangle = 
\frac{1}{2} \,
\sum_{ij} \,  \frac{\langle  B | X_{i}  X_{j} \rangle
P_{i}^L \ P_{j}^L 
}{
(1 - \nu_{i}) (1 - \nu_{j})}  
+ \sum_{i} \, \frac{P_{i}^L}{1 - \nu_{i}}  
\langle  B | \hat{X}^{\dag}_{i} | \bar{\psi}_{1L} \rangle \,
+ B_c .
\end{equation} 
where 
$B_c  
= \langle B | \bar{\psi}_2 \rangle$
describes the X--X and X--X+MP correlations.   
The first term in the above equation describes  the 
familiar Hartree--Fock X--X interactions,  
while the 
second term describes the analogous  
interactions between the exciton $\hat{X}_i$ 
and the 
\{1--$h$/2DEG$^\ast$\} state $|\bar{\psi}_{1L}\rangle$. 
By using Eq. (\ref{B}) for $\hat{B}$
to calculate the overlap 
$\langle G | \hat{B} | X_{i} X_{j} \rangle$ 
and noting  that, in the X--phonon system, 
$\langle G | \hat{B} \hat{X}^{\dag}_{i} | \bar{\psi}_{1L} \rangle=0$, 
we reproduce the results 
obtained in \cite{axt98,axt96} for the undoped system
by using the cumulants. 

We now turn to the 
 equation of motion 
for the correlated X--X amplitude 
$B_c$.  
By projecting the state $\langle B|$ to Eq. (\ref{bar2hole}), 
restricting to the LL0 and LL1 states, and using Eq. (\ref{Ysymm})  
we obtain that 
\begin{align}
\label{eom-Bc}
i \partial_t B_c =&
\langle B | H | \bar{\psi}_2 \rangle 
+ \frac{\langle B | \hat{B}^\dag | G \rangle}{2} 
\left( \frac{P_1^L}{1 - \nu_1} -\frac{ P_0^L}{
1 - \nu_0} \right)^2 \nonumber \\ &
+ \left( \frac{P_1^L}{1 - \nu_1} -\frac{ P_0^L}{
1 - \nu_0} \right) \langle B | \hat{Y}^\dag | \bar{\psi}_{1L} \rangle 
- \left[ \frac{
\langle B |\hat{X}_1^\dag | \bar{\psi}_{1L} \rangle 
}{1 - \nu_1}
- \frac{\langle B |\hat{X}_0^\dag | \bar{\psi}_{1L} \rangle 
}{1 - \nu_0}\right] \bar{P}^L
\end{align} 
By using the Lanczos recursive method \cite{haydock} 
one can generate a basis of 
strongly correlated 2--$h$ states
  similar to 
Section \ref{linear}. 
By describing the effects of the higher Lanczos states 
by introducing a dephasing rate we 
recover the average polarization model results used to describe 
$X$--$X$ correlations and biexciton effects in undoped
semiconductors \cite{chemla-99,schaf-96,shah00}.
Thus the Lanczos
method can be used to derive such a model, which, in the 
case of 2D magnetoexcitons, 
was shown in \cite{shah00} to be a good approximation 
in the case of attractive or  strong repulsive X--X interactions
with appropriate range.  
The Lanczos basis leads to a continued fraction expression of
the X--X amplitude $B_c$, 
similar to the linear response of the 2DEG \cite{kara}, 
and is advantageous 
in the case of strong coupling between the different X momentum 
states, e.g. due to an antibound continuum resonance 
or a bound biexciton  \cite{shah00}.

\section{Magnetoplasmon Dynamics in the coherent regime}
\label{GPA}

In this section we derive from the above theoretical formulation
a generalized average polarization model  that we then 
use to calculate the 
time--integrated three--pulse FWM signal 
in the  coherent
regime. In the latter regime, 
we neglect all incoherent contributions 
due to the photexcited carriers, which are 
described by 
$|\bar{\psi}_{1L} \rangle$.  
As can be seen 
from Eq. (\ref{Mc-def}),
the intraband contribution to the 
FWM signal 
then comes from the coherent MP amplitude 
$\langle G | [ \hat{Y}, \hat{X}_{i}^{\dag}] | \bar{\psi}_0 \rangle$.
Noting the relation 
\begin{equation}   
\langle \bar{\psi}_0  | \hat{M}_n |G \rangle  
= {\mathcal M}_n^*
+ (\delta_{n1} - \delta_{n0}) 
\langle \bar{\psi}_0  | 
H  \Delta \hat{\nu}_n | G \rangle, 
\end{equation} 
obtained by restricting to the first two LLs 
and using Eq. (\ref{commut-nuH}) 
and the eigenvalue equation 
$H | G \rangle=0$, 
we obtain from Eq. (\ref{Mc-def}) 
\begin{equation} 
\langle \hat{M}_n \rangle_c 
= {\mathcal M}_n +  
{\mathcal M}_n^* 
 + (\delta_{n1} - \delta_{n0}) 
\langle \bar{\psi}_0  | 
H  \Delta \hat{\nu}_n | G \rangle. 
\end{equation} 
Similar  to the 
X--X amplitude  $\langle \hat{B} \rangle_c$, 
the amplitude ${\mathcal M}_n(t)$
describes the time evolution 
of the 
 0--$h$ 2DEG$^\ast$ 
state $\hat{M}_n^\dag | G \rangle 
= \hat{X}_{n} | Y \rangle$ discussed 
below.
Using Eqs. (\ref{M-symm}) and 
(\ref{commut-nuH}) we obtain that  
\begin{align}
| M \rangle 
= \hat{M}_1^\dag | G \rangle =
& 
- \hat{M}_0^\dag | G \rangle
+ H \Delta \hat{\nu}_0 | G \rangle
\nonumber\\
&
+ V_{10}  ( 1 - \nu_0) \Delta \hat{\nu}_1 | G \rangle 
-V_{10} ( 1 - \nu_1) \Delta \hat{\nu}_0 | G \rangle
\label{MP-state}   
\end{align} 
which gives the following relation between
${\mathcal M}_1$ and ${\mathcal M}_0$: 
\begin{align}
{\mathcal M}(t)
= {\mathcal M}_1(t) = 
&
-{\mathcal M}_0(t)
+ \langle G | \Delta \hat{\nu}_0 H | 
\bar{\psi}_0 \rangle  \nonumber \\ 
&+ V_{01}  ( 1 - \nu_0) \langle G | \Delta \hat{\nu}_1 | 
\bar{\psi}_0 \rangle
-V_{01} ( 1 - \nu_1) \langle G | \Delta \hat{\nu}_0  |
\bar{\psi}_0 \rangle.
\label{calM-symm}  
\end{align} 
In the case of the spin--polarized 2DEG (QHE ferromagnet), 
only spin--$\uparrow$ states are populated 
in the ground state and therefore 
$ \Delta \hat{\nu}_n | G \rangle =0$
(see Eq. (\ref{Dnu})).
This is also the case if 
we neglect any fluctuations 
in the filling factor of the spin--$\downarrow$ 
LLs in the ground state.
We then obtain by using Eq. (\ref{MP-state}) 
that 
\begin{equation} 
\langle G | \hat{M}_0  \hat{M}^\dag_0 | G \rangle 
= - \langle G | \hat{M}_0  \hat{M}^\dag_1| G \rangle 
= \langle G | \hat{M}_1  \hat{M}^\dag_1 | G \rangle 
= - \langle G | \hat{M}_0  \hat{M}^\dag_1| G \rangle= W_M
\end{equation} 
and, from Eq. (\ref{calM-symm}), we obtain the relation 
\begin{equation} 
{\mathcal M}(t)={\mathcal M}_1(t)= -{\mathcal M}_0(t).
\end{equation} 
An  explicit expression for the state  $| M \rangle$
can be obtained by acting on the ground state
with the intraband operator 
Eq. (\ref{M-2q}). 
Noting that the 
valence band is full in the ground state, 
the states that contribute to $|M \rangle$ 
are the MP states $\hat{\rho}^e_{{\bf q}nm\sigma} | G \rangle$ 
and the states $ \hat{\rho}^e_{-{\bf q} 01 \downarrow} 
\hat{\rho}^e_{{\bf q} nm \sigma}| G \rangle$
and  $ \hat{\rho}^e_{-{\bf q} 10 \downarrow} 
\hat{\rho}^e_{{\bf q} nm \sigma}| G \rangle$ 
that
vanish in the case of a spin--polarized ground state. 
As already known 
from inelastic light scattering experiments, 
the  MP states $\hat{\rho}^e_{{\bf q}nm\sigma} | G \rangle$ 
can be photoexcited in the real system 
due to the residual disorder, inhomogeneity,  
and valence band mixing \cite{aron-rev,aron-92,raman,marmorkos}.
For exaqmple, MPs with momentum 
much larger than the photon momentum,  
close to the 
maximum of the MP dispersion or 
close to the magnetoroton dispersion minimum 
 $q \sim 1/l$, 
have been observed in such experiments
 \cite{aron-rev,aron-92,raman,marmorkos}.
For the filling 
factors and photoexcitation conditions of interest here, 
the  LL0 $\to$ LL1 MP will 
dominate, due to its longer lifetime 
and resonant contribution. 

To evaluate the equation of motion for 
${\mathcal M}(t)$ we also need the 
state $H |M \rangle$, which can be obtained 
by using the Lanczos approach
\begin{equation} 
H | M \rangle = 
\Omega_M | M \rangle + |\bar{M}\rangle \ , \ 
\Omega_M  = \frac{\langle M | H | M \rangle}{\langle M | M \rangle}
\ , \ \langle \bar{M} | M \rangle=0. \label{HonM}  
\end{equation} 
Neglecting $|\bar{M}\rangle$ corresponds to the single mode approximation 
\cite{QHE1,QHE2,oji}. 
The above correlated  basis set leads to a 
a continued fraction expansion for 
${\mathcal M}(\omega)$ similar to the linear polarization 
calculation of 
Section \ref{linear} \cite{kara},  
the correlated X--X amplitude calculation 
outlined in Section \ref{XX} \cite{shah00},  
and the 2DEG dynamical structure factor calculations 
disucssed e.g. in \cite{QHE1,QHE2}. 
 
Neglecting 
all incoherent contributions and using the above relations  
we  obtain from Eq. (\ref{bar0hole}) 
or  Eq. (\ref{eom-Mcorr}) 
the equation of motion 
\begin{equation} \label{M-eom}
i \partial_t {\mathcal M} =
(\Omega_{M} - i \gamma_M) {\mathcal M}
+W_M
\left( \frac{ P_1^{L*}}{1-\nu_1} 
 - \frac{ P_0^{L*}}{1 - \nu_0} \right) 
\left( \frac{ P_1^{L}}{1-\nu_1} 
 - \frac{ P_0^{L}}{1 - \nu_0} \right),
\end{equation}
where $\gamma_M$ describes the damping of the 2DEG excitation 
$| M \rangle$.  
Neglecting the incoherent 
non-linear contributions to Eq. (\ref{Y-c}) 
we also 
that 
\begin{equation} 
i \partial_t \langle \hat{Y} \rangle_c 
= ( \bar{\Omega} - i \gamma) 
\langle \hat{Y} \rangle_c  + 
W \left( \frac{ P_1}{1 - \nu_1} 
- \frac{P_0}{1 - \nu_0}
\right),
\end{equation} 
which describes the dephasing of the non-linear 
polarization similar to the case of the linear absorption. 
Substituting the above results into 
the equation of motion Eq. (\ref{eom-Pi})
 we obtain for the 
third order non-linear polarization  
in the coherent regime 
\begin{align}
i \partial_t & P_n
-
(\Omega_n - i\Gamma_n) P_n +  V_{nn'} ( 1- \nu_n) P_{n'}
=
\frac{ 2 \mu E(t) P_{n}^{*} P_{n}}{1 - \nu_n} 
+ (\delta_{n1}-\delta_{n0}) \times
\nonumber \\
&
\left[
V_{XX} 
  \left( \frac{P_1 P_0 P_1^{*}}{1-\nu_1} 
 - \frac{ P_1 P_0P_0^{*}}{1 - \nu_0} \right)
+ \left( {\mathcal M}\!+\!{\mathcal M}^* \right)  
\left( \frac{ P_1}{1-\nu_1} 
-\frac{ P_0}{1 - \nu_0} \right) 
+ 
\langle \hat{Y} \rangle_c\right]
\label{Pmodel} 
\end{align}
where $n,n'=0,1$, 
$n \ne n'$, 
The first term on the rhs of the above equation 
describes the 
Pauli blocking effects due to the 
coherent exciton density $P_nP_n^{*}$. The second term describes the 
mean field X--X interactions, 
where 
$V_{XX}= \langle B | 
X_1 X_0 \rangle$ 
is the only finite contribution to 
Eq. (\ref{2X}).
The above terms reproduce the results of \cite{staff90}
if we restrict to the first two LLs. 
The third term in Eq. (\ref{Pmodel}) 
describes the effects of the 
photoexcitated  MP coherence 
${\mathcal M}$, whose time evolution is determined by the 
MP dynamics, while the last term describes the non-Markovian 
dephasing of the X polarization, which is  governed by the 
time evolution of the X+MP states $| Y \rangle$.  
Using the above equations, we calculate the 
time--integrated FWM signal 
\begin{equation}\label{TIFWM}
S(\dt{2},\dt{3})=
\int_{-\infty}^{+\infty}dt\:|P_0(t)+P_1(t)|^2
\end{equation}
in the ${\bf k}_s
= {\bf k}_1+{\bf k}_2-{\bf k}_3$ direction, 
assuming right--circularly polarized optical excitation of the form
$E(t) =
e^{i{\bf k}_1\cdot{\bf r}}\mathcal{E}_p(t) +
e^{i{\bf k}_2\cdot{\bf r}}\mathcal{E}_p(t + \dt{2})+
e^{i{\bf k}_3\cdot{\bf r}}\mathcal{E}_p(t + \dt{3})$,
where ${\mathcal E_p}(t)$ is the Gaussian optical pulse.
In the next section we discuss the results of this 
simple model that describes 
only coherent effects. 
The role of the  MP in the population relaxation 
and the role of MP--assisted 
inter--LL coherences among the photoexcited carriers 
will be discussed elsewhere. 

%%%%%%%%%%%%%%%%%%%%%%%%%%%%%%%%%%%% NUMERICAL RESULTS

\section{Three--pulse time integrated four--wave--mixing signal} 
\label{results}

In this section we present 
the results of our numerical and analytical calculations 
based on the model of Section~\ref{GPA} and using 
 parameters of Fig. \ref{LA}, 
obtained by fitting
the experimental linear 
absorption as discussed in \cite{kara,from-02-prb}. 
Below we analyze the 
time dependence of the TI--FWM signal 
as a function of two time delays: 
$\dt{3}$, 
the delay between pulses 1 and 3
also accessible in two--pulse FWM experiments, 
and $\dt{2}$,  
the time delay between 
pulses 1 and 2.
 We find that  $\dt{2}$ can 
give direct information on the MP dynamics that is not accessible 
as function of  $\dt{3}$, 
where the time--dependence is similar to 
the results of \cite{kara}. 
We study below the FWM signal 
that comes from the three different non-linearities in Eq. (\ref{Pmodel}), 
i.e. the PSF, X--X interactions, and MP coherence.   
Furthermore, we show that 
the correlation effects can be controlled by 
varying 
the central photoexcitation frequency 
from LL1 toward LL0. 
Important is the fact that, as already  seen 
in the linear absorption spectra (Fig. \ref{LA})  and 
discussed in \cite{from-02-prl,kara,from-02-prb}, 
the LL1 magnetoexciton decays much faster than  
the LL0 magnetoexciton  due to the 
X+MP dynamics described by $\bar{P}$. 
For simplicity, when we discuss the numerical 
results below, we refer to exciton dephasing rates 
$\Gamma_0$ and $\Gamma_1$ 
of $P_0$ and $P_1$ respectively, 
with $\Gamma_1 \gg \Gamma_0$. 
In the experiment \cite{dani}, the LL1 dephasing
was measured to be of the order of a few hundreds of 
femtoseconds, while the LL0 dephasing time 
was of the order of a few picoseconds. 
% 
% FIG.2: PSF
\begin{figure}[t]
\begin{center}
\includegraphics*[width=13.5cm]{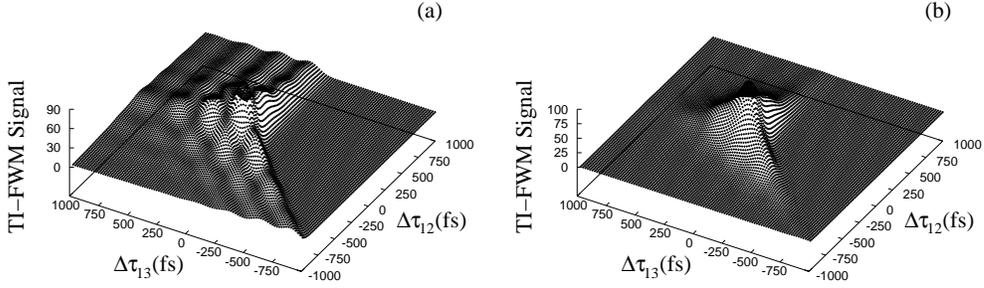}
\caption{\label{PSF}
PSF contribution to the TI--FWM signal when exciting
(a) LL0 and LL1 equally
(b) mostly LL1  (arbitrary units), 
with parameters as in Fig. (\ref{LA}).
}
\end{center}
\end{figure}
Fig.~\ref{PSF}
shows the PSF contribution to the TI--FWM signal, 
obtained by only retaining the first term on the rhs of Eq. (\ref{Pmodel}), 
whose time dependence is determined
by the dephasing of the LL0 and LL1 magnetoexcitons. 
The PSF signal 
is maximum at $\dt{2}=\dt{3}= 0$ and 
peaked along 
the directions $\dt{3}>0, \dt{2}=0$,
$\dt{2}>0, \dt{3}=0$ and 
$\dt{3}=\dt{2}<0$. 
Fig.~\ref{PSF}
compares the PSF contribution for 
optical 
excitation centered at LL1
(Fig.~\ref{PSF}(b))
and between the two LL peaks (Fig.~\ref{PSF}(a)).
Fig.~\ref{PSF} demonstrates a strong dependence of the PSF
temporal behavior on the central photoexcitation frequency, 
which is due to the 
 much faster  dephasing 
of $P_1$ as compared to 
 $P_0$ and the fact that the PSF contribution 
to Eq. (\ref{Pmodel}) is directly proportional to 
the optical pulse. 
When the excitation frequency is centered on
LL1, 
the PSF signal is dominated by the LL1 density 
contribution, 
while the LL0 density is suppressed. 
In this case, the PSF signal 
decays very fast, as 
$\sim 2\Gamma_1(\dt{2}+\dt{3})$ 
in the $\dt{2}>0$, $\dt{3}>0$ region and as 
$\sim 2\Gamma_1(\dt{3}-2\dt{2})$ in the $\dt{3}>\dt{2}$, 
$\dt{2}<0$ region. On the other hand, 
when both LLs are equally excited, 
the long--lived LL0 exciton contribution
dominates over the short--lived LL1 exciton contribution  
for sufficiently long time delays, 
and thus the signal  decays more slowly as compared to 
the case of LL1 photoexcitation since $\Gamma_0 \ll \Gamma_1$.
Moreover, beatings occur 
between the LL0 and LL1 contributions, 
with frequency 
$\sim \Omega_1-\Omega_0$, 
the difference 
in the energies between the two X peaks,
and decay rate $\sim \Gamma_0+\Gamma_1$.
We conclude that the PSF contribution to the TI--FWM experimental signal 
is characterized by a  strong dependence
on the central photoexcitation frequency, 
due to its proportionality to the optical pulse 
and the large difference in the dephasing of the two LL excitons 
due to the X+MP states. 

\begin{figure}[t]
\begin{center}
\includegraphics*[width=13.5cm]{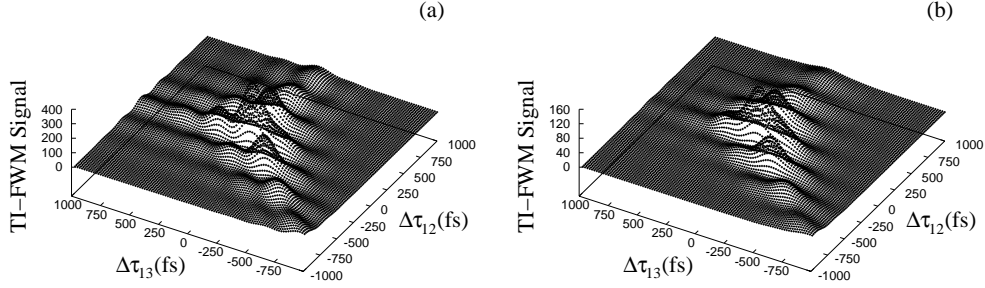}
\caption{\label{XX-fig}
TI--FWM signal due to the $X$--$X$ interactions alone when exciting
(a) LL0 and LL1 equally, 
and 
(b) mostly LL1 (arbitrary units), with parameters 
as in Fig. (\ref{LA}). 
}
\end{center}
\end{figure}

Fig.~\ref{XX-fig} shows the FWM signal due to 
the X--X interactions alone, 
obtained by only retaining the second term on the rhs of  Eq. (\ref{Pmodel}).  
The overall temporal profile is similar to the PSF signal, 
with strong oscillations 
as function of  $\dt{2}$
with frequency 
$\Omega_1-\Omega_0$.
By shifting the central photoexcitation 
frequency 
towards LL0,
the XX signal along the $\dt{3}>0, \dt{2}=0$
direction and the corresponding beatings are enhanced. 
%
% FIG.3: X-X interactions

\begin{figure}[t]
\begin{center}
\includegraphics*[width=13.5cm]{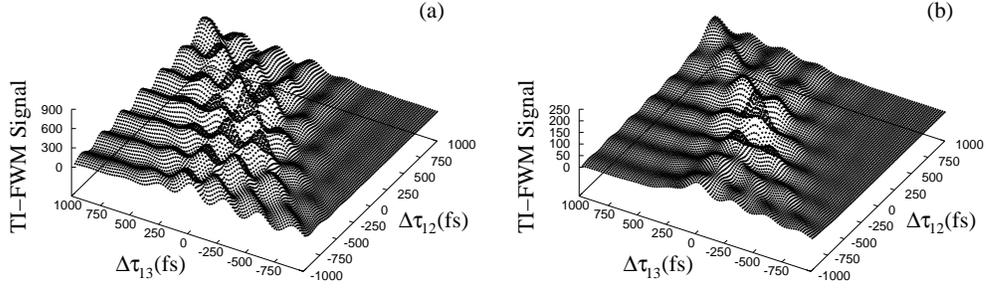}
\caption{\label{MP-fig}
Coherent magnetoplasmon contribution to the TI--FWM signal when exciting
(a) LL0 and LL1 equally, and 
(b) mostly LL1  (arbitrary units), 
with parameters as in Fig. (\ref{LA}). 
The MP correlations 
create a signal in the 
$\dt{2}=\dt{3}>0$ and $\dt{2}<0=\dt{3}$ directions
that decays with the MP dephasing time $\gamma_M=0.1\:meV$. 
}
\end{center}
\end{figure}

The coherent MP  contribution,
obtained from the third term on the rhs of 
Eq. (\ref{Pmodel}), is shown in Fig.~\ref{MP-fig}. 
 The new feature 
that distinguishes the MP contribution from the rest 
in the case of a long MP dephasing time 
is 
the strong signal in the directions $\dt{2}<0, 
\dt{3}=0$ and $\dt{2}=\dt{3}>0$, 
which is due to the 
photoexcitation 
of a MP with energy $\Omega_M$ 
comparable to the difference between the 
LL1 and LL0 exciton energies.
This resonance between the MP energy and 
exciton splitting enhances the MP--induced signal, 
and leads to an optically--induced 
 time--dependent hybridization of the  
LL0 and LL1 magnetoexcitons, which is  
governed by the MP dynamics described by 
${ \mathcal M}$ and comes from the absorption 
of a photoexcited MP that couples the LL1  
to the LL0 exciton as described by $P_1 {\mathcal M}^*$. 
The above signal is mainly due to the 
resonant, 
$\propto  P_1 P_0^*$, 
contribution to the equation of motion 
Eq. (\ref{M-eom}), 
which describes a photoexcited inter--LL coherence that 
subsequently scatters into a MP 2DEG coherence 
via the X--2DEG interactions. 
Moreover, Fig.~\ref{MP-fig} shows 
strong oscillations as function of both time delays,  
which are mainly due to the 
non-resonant  contribution 
to ${ \mathcal M}$,  $\propto P_0 P_0^*$ in Eq. (\ref{M-eom}), 
and are thus of quantum 
kinetic origin. 
By shifting the photoexcitation frequency 
from LL1 toward LL0,  
$P_0$ 
increases and thus the amplitude of these beatings,  
as well as the overall strength of the MP FWM signal 
is enhanced 
(see Fig. \ref{MP-fig}(a)).
Of particular interest for studying the 
MP dynamics is 
the strong signal and oscillations 
in the directions $\dt{2}<\dt{3}=0$ and $\dt{2}=\dt{3}>0$. 
The FWM signal in the above directions 
decays overall  
as $\sim 2\gamma_M$, the MP dephasing rate, 
and $\sim 4 \Gamma_0$ 
and displays oscillations
with frequency
$\sim \Omega_M$, the MP energy, 
that decay with a rate of  $\sim 2\Gamma_0+\gamma_M$.
Therefore, the above signal can be used to 
extract from a three--pulse 
TI--FWM 
experiment the MP dynamics and dephasing rate.
In contrast, the rest of the signal,  
along directions $\dt{2}>0, \dt{3}=0$ or 
$\dt{3}>0, \dt{2}=0$ or $\dt{2}=\dt{3} <0$ 
decays overall as $\sim 2 \Gamma_0$ and shows oscillations 
with frequency $\sim \Omega_1 - \Omega_0$ 
that decay with a rate $\sim \Gamma_0 + \Gamma_1$.
Finally, along the direction $\dt{3}<0, \dt{2}=0$, 
the signal decays overall with a rate $\sim 4 \Gamma_0$ 
and shows oscillations with frequency 
$\sim \Omega_1 - \Omega_0$ 
that decay with a rate $\sim 3 \Gamma_0 + \Gamma_1$.
%
% FIG.4: MP

\begin{figure}
\begin{center}
\includegraphics*[width = 10cm]{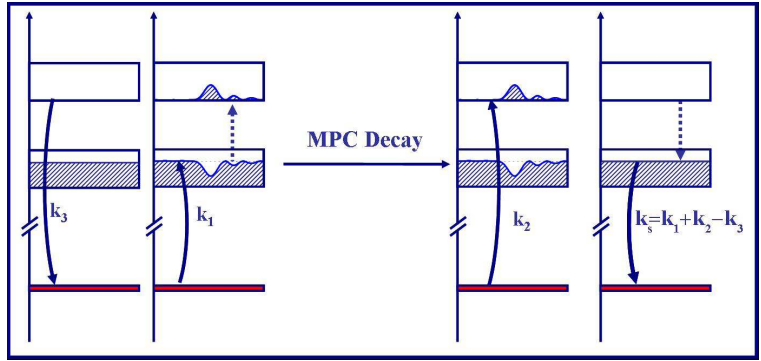}
\caption{\label{schematic}
Non-Linear process that gives
the MP  FWM signal.
This corresponds to an inelastic Raman process, assisted 
by the interactions between the photoexcited 
Xs and the 2DEG. 
}
\end{center}
\end{figure}

The main physical mechanism that gives the 
signal in the $\dt{2}<0, \dt{3}=0$ direction is shown schematically  
in Fig. \ref{schematic} and  may be summarized as follows.  
%
% FIG.6: SCHEMATIC ILLUSTRATION
The ${\bf k}_3$ and ${\bf k}_1$ pulses arrive at the same 
time and photoexcite an inter--LL coherence that 
subsequently scatters into a  
MP polarization due to the X--2DEG interactions. 
The ${\bf k}_2$ pulse comes in later and creates an X polarization  
that scatters off the above decaying MP and 
 gives  a signal in the 
${\bf k}_1+{\bf k}_2-{\bf k}_3$ direction. 
In particular, 
the MP scatters 
with the photoexcited X into an X state
that then recombines to return the system in the ground state. 
It is interesting to note the
similarity of the above process and the familiar one of coherent
antiStokes Raman scattering \cite{levenson-82} that
involves phonons.   
Note here that, in the case of a spin--polarized 
2DEG, the Raman process of excitation and de-excitation 
of the system with right--circularly 
polarized light is negligible due the absence of 
spin--$\downarrow$ electrons in the ground state. 
A MP excitation
can however be excited 
due to the scattering of  the photoexcited spin--$\downarrow$ X 
with the  spin--$\uparrow$  2DEG. 
Obviously, if the MP dephases before the ${\bf k}_2$ pulse arrives, 
there will be no 
FWM signal for this particular sequence of time delays, 
and therefore the above direction can be used to 
extract  the MP dephasing dynamics from the experimental results. 
Furthermore, as discussed above, the   $\sim \Omega_M$ oscillations 
for $\dt{2}=\dt{3}>0$ or 
 $\dt{2}<0, \dt{3}=0$ directions are 
of quantum kinetic origin and their 
decay gives important information on the 
MP coherent dynamics during  time scales comparable to the 
MP  period.  
%
% FIG.5: Full Signal
\begin{figure}[t]
\begin{center}
\includegraphics*[width=13.5cm]{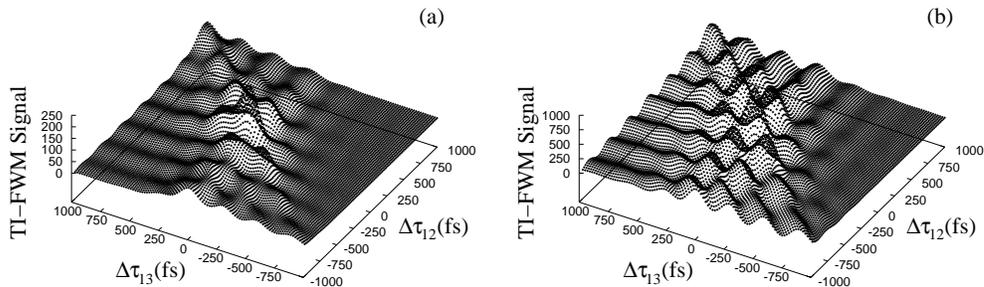}
\caption{\label{signal}
TI--FWM signal with PSF + MP contributions when exciting
(a)  LL0 and LL1 equally
(b) mostly LL1 (arbitrary units) 
The parameters are as in Fig. (\ref{LA}), 
$W_M = W =V_{XX} = 3\: meV$, $\gamma_M = 0.1\: meV$. 
}
\end{center}
\end{figure}
Fig.~\ref{signal} shows the TI signal 
when both the MP and  PSF contributions are included.  
This figure shows a rich  time--dependence, which is dominated by the MP 
contribution
and can be used as a guide to extract from 
a three--pulse FWM experiment
new information 
about the dephasing dynamics of the $X_1$ and $X_0$ excitons,  
as well as the MP dynamics 
during time scales comparable to the 
MP period. 
A comparison between the theory and such an experiment \cite{dani}
as well as the role of the incoherent effects neglected in the above model 
will be presented elsewhere. 

\section{Conclusions} 
\label{concl} 

In summary, we
discussed a recent 
theory \cite{kara,per-03,per-dan-ssc} 
that provides a unified description
of the ultrafast 
non-linear optical response
of magnetoexcitons  
in both doped and undoped 
semiconductors, including systems 
with a  strongly correlated 
many--electron ground state, 
such as the 
2DEG in the QHE regime. 
We discussed a method for describing the 
interaction contributions 
to the density matrix equation of motion, which 
gives the third--order non-linear  polarization measured in 
transient wave mixing and pump--probe 
experiments and established the connection with the 
DCTS treatment of X--phonon and X--X interactions 
in the undoped system \cite{rossi,axt98,axt96}. 
Using a decomposition of the photoexcited many--body wavefunction 
into correlated/uncorrelated and excitonic/incoherent 
contributions, we obtained a factorization 
scheme for the density matrix that 
separates out the correlated from the uncorrelated contributions. 
Importantly, our method 
applies 
to systems with a strongly correlated ground state, 
populated by a 2DEG here, where previous factorization 
schemes, based on 
a Hartree--Fock ground state, need to be extended. 
Our formulation may be used to study the 
role of exciton--2DEG and exciton--exciton correlations
on the ultrafast non-linear optical response 
 and the interplay 
between coherent and incoherent effects.
Our expansion in terms of the optical 
field is valid for sufficiently  short pulses 
and/or weak excitation conditions, 
where the correlations are most pronounced. 
We describe the role 
of the long--lived collective excitations 
of a strongly correlated cold electron gas, which is  
 present prior to the optical excitation.
We also presented a number of results that describe the 
Pauli exchange and interaction effects between 
excitons and magnetoplasmon collective excitations. 

We applied
our theory to the case of the photoexcitation 
of the 2DEG 
with three  right--circularly polarized 
optical pulses.
Our numerical solution 
for 
the time--integrated three--pulse FWM signal 
demonstrates a  signature 
of  the collective 2DEG excitation 
dynamics 
that, for 
long MP dephasing 
time, 
 can be distinguished from 
the Pauli blocking and exciton--exciton 
interaction contributions. 
In the case of interest here, 
the relevant 2DEG 
collective excitations  
are the long--lived inter--LL magnetoplasmons, 
which  dress the 
photoexcited magnetoexcitons and lead to polaronic--like 
effects
and  strong non-Markovian dephasing and quantum kinetic effects.
We showed that such effects dominate 
the time delay and frequency 
dependence  of the transient FWM
signal in the coherent regime  for time scales 
comparable to the inverse MP frequency. 
FWM spectroscopy 
using femtosecond optical pulses provides both the 
time and the frequency resolution 
necessary to access a new regime 
of 2DEG and QHE physics.
Our theory allows us to study 
the experimental signatures 
of the 2DEG 
quantum 
dynamics.
We  predicted,  in particular, 
strong
non-Markovian effects, leading to 
an asymmetric LL1 exciton lineshape 
and to strong oscillations as function of time delay
 of quantum kinetic origin
that are governed by the MP dynamics. 
We also found an optically--induced hybridization of the LL0 and LL1 excitons, 
with the time--dependent coupling determined by the MP dynamics. 
A comparison 
of our results with recent experiments
\cite{dani} is currently in progress and will be presented elsewhere. 

The above correlation--induced non-Markovian  dynamics 
can be controlled  
by tuning the central frequency of the optical 
excitation between the two lowest LLs, 
which changes the coherent admixture of the two 
MP--dressed magnetoexcitons. 
FWM experiments
using a sequence of 
optical pulses with different polarizations 
and central frequencies 
provide new ways 
for accessing 
the very early dynamics 
of the strongly correlated  2DEG, during time scales 
shorter than the duration of the interaction processes. 
This opens up a new window into non-equilibrium 
Quantum Hall  effect
physics and collective effects that are just now starting to be explored.
In particular, interesting regimes include {\bf (1)} filling factors
away from $\nu=1$, where skyrmions \cite{QHE2} become important
in one--sided modulation doped quantum wells 
and are expected to  dominate the magnetoexciton dephasing, 
and {\bf (2)} fractional filling factors,  
where intra--LL collective excitations 
will dress the magnetoexciton leading to polaronic effects \cite{apalkov}. 

%%%% ACKNOWLEDGEMENTS  %%%%%

\section*{ Acknowledgements}

This work is the result of a long and close collaboration with Daniel Chemla. 
We also thank K. Dani, J. Tignon, N. Fromer, and A. Karathanos  for 
their collaboration on various aspects of this project,
and S. Cundiff for useful discussions. 
This work 
was supported by the EU Research Training Network  HYTEC
 (HPRN-CT-2002-00315) and  
by the U.S. Department of Energy  under grant 
No. DE-FG02-01ER45916. 

%%%  APPENDICES %%%
% The Appendices part is started with the command \appendix;
% appendix sections are then done as normal sections
% \appendix
% \section{}
% \label{}

%%%  BIBLIOGRAPHY %%%
% \bibitem{label}
% Text of bibliographic item
% notes:
% \bibitem{label} \note
% subbibitems:
% \begin{subbibitems}{label}
% \bibitem{label1}
% \bibitem{label2}
% If there is a note, it should come last:
% \bibitem{label3} \note
% \end{subbibitems}

%%%%%%%%%%%%%%%%%%%%%%%%%%%%%%%%%%%%%%%%%%%%%%%%%%%%%%%%%%%%%%%%%%%%%%%%%%

\end{document}